\documentclass{aa}
\usepackage[varg]{txfonts}
\usepackage{graphicx}
\usepackage[version=3]{mhchem} 
\usepackage{longtable}
\bibpunct{(}{)}{;}{a}{}{,} 
\usepackage{color}

\begin{document} 

\title{Tracing the cold and warm physico-chemical structure of deeply embedded protostars: IRAS~16293-2422 versus VLA~1623-2417}

\author{N. M. Murillo\inst{1} \and E. F. van Dishoeck\inst{1,2} \and M. H. D. van der Wiel\inst{3,4} \and J. K. J{\o}rgensen\inst{3} \and M. N. Drozdovskaya\inst{1,5} \and H. Calcutt\inst{3} \and D. Harsono\inst{1}}

\institute{Leiden Observatory, Leiden University, P.O. Box 9513, 2300 RA, Leiden, the Netherlands \\ \email{nmurillo@strw.leidenuniv.nl}
        \and Max-Planck-Institut f\"{u}r extraterrestrische Physik, Giessenbachstra\ss e 1, 85748, Garching bei M\"{u}nchen, Germany
		\and Centre for Star and Planet Formation, Niels Bohr Institute \& Natural History Museum of Denmark, University of Copenhagen, {\O}ster Voldgade 5–7, DK-1350 Copenhagen K., Denmark
		\and ASTRON, the Netherlands Institute for Radio Astronomy, Postbus 2, 7990 AA Dwingeloo, The Netherlands
		\and Center for Space and Habitability (CSH), University of Bern, Sidlerstrasse 5, CH-3012 Bern, Switzerland}

\abstract
{Much attention has been placed on the dust distribution in protostellar envelopes, but there are still many unanswered questions regarding the physico-chemical structure of the gas.} 
{Our aim is to start identifying the factors that determine the chemical structure of protostellar regions, by studying and comparing low-mass embedded systems in key molecular tracers.}
{The cold and warm chemical structures of two embedded Class 0 systems, IRAS~16293-2422 and VLA~1623-2417 is characterized through interferometric observations. \ce{DCO+}, \ce{N2H+} and \ce{N2D+} are used to trace the spatial distribution and physics of the cold regions of the envelope, while \ce{c-C3H2} and \ce{C2H} from models of the chemistry are expected to trace the warm (UV-irradiated) regions.}
{The two sources show a number of striking similarities and differences. \ce{DCO+} consistently traces the cold material at the disk-envelope interface, where gas and dust temperatures are lowered due to disk shadowing. \ce{N2H+} and \ce{N2D+}, also tracing cold gas, show low abundances towards VLA~1623$-$2417, but for IRAS~16293$-$2422, the distribution of \ce{N2D+} is consistent with the same chemical models that reproduce \ce{DCO+}. \ce{c-C3H2} and \ce{C2H} show different spatial distributions for the two systems. For IRAS~16293$-$2422, \ce{c-C3H2} traces the outflow cavity wall, while \ce{C2H} is found in the envelope material but not the outflow cavity wall. In contrast, toward VLA~1623$-$2417 both molecules trace the outflow cavity wall. Finally, hot core molecules are abundantly observed toward IRAS~16293$-$2422 but not toward VLA~1623$-$2417.}
{We identify temperature as one of the key factors in determining the chemical structure of protostars as seen in gaseous molecules. More luminous protostars, such as IRAS~16293-2422, will have chemical complexity out to larger distances than colder protostars, such as VLA~1623-2417. Additionally, disks in the embedded phase have a crucial role in controlling both the gas and dust temperature of the envelope, and consequently the chemical structure.}
   
\keywords{astrochemistry - stars: formation - stars: low-mass - ISM: individual objects: IRAS~16293-2422 and VLA1623-2417 - methods: observational - techniques: interferometric}

\titlerunning{Comparing the chemical structure of IRAS~16293-2422 and VLA~1623-2417}
\authorrunning{Murillo et al.}
   
\maketitle

\section{Introduction}
While there is a well-established outline of the physical evolution of protostellar systems \citep{evans1999,dunham2014,li2014,reipurth2014}, there are still many questions regarding the physico-chemical structure of these systems.
Several studies point out the chemical richness and diversity of young embedded protostars, most notably in the Class 0 stage, ranging from simple molecules to carbon chains and complex organics \citep[see reviews by][]{herbst2009,caselli2012,sakai2013}. 
In contrast, some other protostellar systems show much less chemical complexity \citep[e.g.,][]{jorgensen2005h2co,maret2006,oberg2014,fayolle2015,lindberg2014a,lindberg2016,lindberg2017,bergner2017}, a situation made more extreme when some starless cores have stronger molecular line emissions than the already formed protostars (e.g., \citealt{bergman2011,bacmann2012,friesen2014}).
It is interesting to explore the chemical structure and evolution of early stage protostars and what physical quantities dictate the resulting chemical structure as observed in the gas phase.

The chemical fingerprint generated in the early embedded stages of
star formation may be transmitted to the later stages and
eventually the protoplanetary disk, where planets and comets are formed \citep[e.g.,][]{aikawa1999a,aikawa1999b,visser2009,visser2011,hincelin2013,drozdovskaya2014,willacy2015,yoneda2016}.
Which factors then generate a protostellar system's fingerprint?
Protostellar cores may inherit
their chemical composition from the parent clouds that eventually collapse to form protostars \citep[e.g.,][]{visser2009,visser2011,aikawa2012,furuya2012,tassis2012,hincelin2016}.
It would then seem likely that protostars from the same parent cloud would have a similar chemical composition.
However, this would require the cloud to have a homogeneous composition, which is not always the case \citep{bergman2011}.
Instead, other mechanisms could alter the chemical fingerprint.
Turbulence and large-scale motions could stir the gas and dust of the cloud core around, moving material from the outer region of the core closer to the warmer regions of the system, kick-starting chemical reactions and producing enhancements of selected species.
Formation of more complex chemical species likely occurs through grain-surface reactions (i.e., on ice and dust surfaces) instead of in the gas-phase, and such reactions proceed faster at higher dust temperatures which increases the mobility of radicals \citep{garrod2006}. If material near outflow cavities is warmer than elsewhere in the envelope, 
this could generate pockets of chemically rich ices that, once heated above the sublimation temperature, would be released into the gas-phase \citep{maria2015}.
Moreover, UV radiation can photodissociate CO and create free atomic carbon that leads to efficient formation of carbon-bearing molecules. UV irradiation together with age and variations in accretion rates would also produce different outcomes, even with the same initial ingredients.
In addition, simple warm-chemistry molecules can be the precursors to
more complex molecules \citep{sakai2013}.

The physical evolution of the individual protostars, e.g., the collapse time and structure, will also impact the chemical fingerprint.
An important consideration regarding the physical structure is that disks may have formed already in the early stages, as shown by recent observations
(e.g., \citealt{tobin2012,murillo2013,harsono2014,lindberg2014,codella2014,yen2017}). Not only do disks provide a high density long-lived reservoir preventing molecules from falling into the star, but they also affect the thermal structure
of their surroundings. 
Thus the disk-envelope interface and the envelope itself must be
studied \citep{murillo2015,persson2016}.  The disk-envelope interface
and the outer envelope of embedded systems are traced by
cold-chemistry molecules, since these regions are usually shielded
from heating by the central protostar
\citep{ewine1995,jorgensen2004,jorgensen2005,sakai2014N,murillo2015}. Through the study of 
molecules sensitive to temperature, we can then understand the structure of embedded
protostellar systems.

Aiming to explore the chemical evolution of the earliest embedded
protostellar systems, i.e. Class 0, we compare two systems from $\rho$
Ophiuchus ($d \sim$ 120 pc, \citealt{loinard2008}), IRAS~16293-2422 and
VLA~1623-2417, separated by a projected distance of 2.8 pc.
Most previous studies were based on single-dish studies. 
The advent of the Atacama Large Millimeter/submillimeter Array (ALMA)
now allows chemical studies on 100~AU scales that
spatially resolve the different physical components of the system.

IRAS~16293$-$2422 (hereafter IRAS~16293) is a widely studied
multiple system, located in L1689N, with a complicated outflow
structure being driven by source A
\citep{stark2004,yeh2008,loinard2013,kristensen2013,girart2014}.  IRAS~16293~A  and B, separated by
about 620~AU, have different inclination angles, with A's disk-like
structure being inclined and B orientated face-on with respect to the line of
sight \citep{pineda2012,jorgensen2016}.  Due to the different
inclination angles, it is difficult to determine whether these systems
are at the same evolutionary stage or not \citep{murillo2016}.  Both
components are chemically rich but show differences in structure
\citep{bottinelli2004,bisschop2008,jorgensen2011}.

VLA~1623$-$2417 (hereafter VLA~1623) is a triple protostellar
system, located in L1688 ($\rho$ Oph A), mostly studied for its
prominent outflow in the region \citep{andre1990,alessio2006}.  
The three components of the system, VLA~1623 A, B and W are separated by 
132 and 1200~AU, respectively, have similar inclination angles,
and have also been found to be at different evolutionary stages
\citep{murillo2013a,murillo2013}.  VLA~1623 has been
shown to be largely line poor in single-dish studies
\citep{garay2002,jorgensen2004,bergman2011,friesen2014}.

In this paper, we present observations of \ce{DCO+}, \ce{N2H+},
\ce{N2D+}, \ce{c-C3H2} and \ce{C2H} towards IRAS~16293 and VLA~1623, using
ALMA, the Submillimeter
Array (SMA) and the Atacama Pathfinder EXperiment (APEX; \citealt{gusten2006}). \ce{DCO+}, \ce{N2H+} and \ce{N2D+}
are known to be good tracers of cold gas where CO is frozen out.
\ce{c-C3H2} and \ce{C2H} are usually seen in photon-dominated regions
(PDRs) such as the Orion Bar \citep{pety2007,wiel2009,nagy2015} and
the Horsehead Nebula \citep{cuadrado2015,guzman2015}, with both
species located at the irradiated, and thus warmer, edge of these
regions.  \ce{c-C3H2} and \ce{C2H} could thus be expected to trace the
(UV-irradiated) outflow cavity walls, although both species have also
been found just outside the disk-envelope interface
\citep{sakai2014}. Besides
mapping their distributions, multiple lines from a single molecule can
also be used to trace physical conditions such as temperature and
density \citep{ewine1993,evans1999,vandertak2007,shirley2015} and the current dataset
allows this to be done for several species.

Details of the observations with ALMA, SMA and APEX
are described in Sect.~\ref{sec:obs}.
Section~\ref{sec:res} describes the spatial distribution
of each molecule for both systems.  The observations are compared to
chemical models and physical parameters are derived in
Sect.~\ref{sec:analysis}.  Sections~\ref{sec:dis} and \ref{sec:conc}
compare both systems studied here with other objects found in
literature and place the results of our work in context.

\begin{figure*}
	\includegraphics[width=0.5\textwidth]{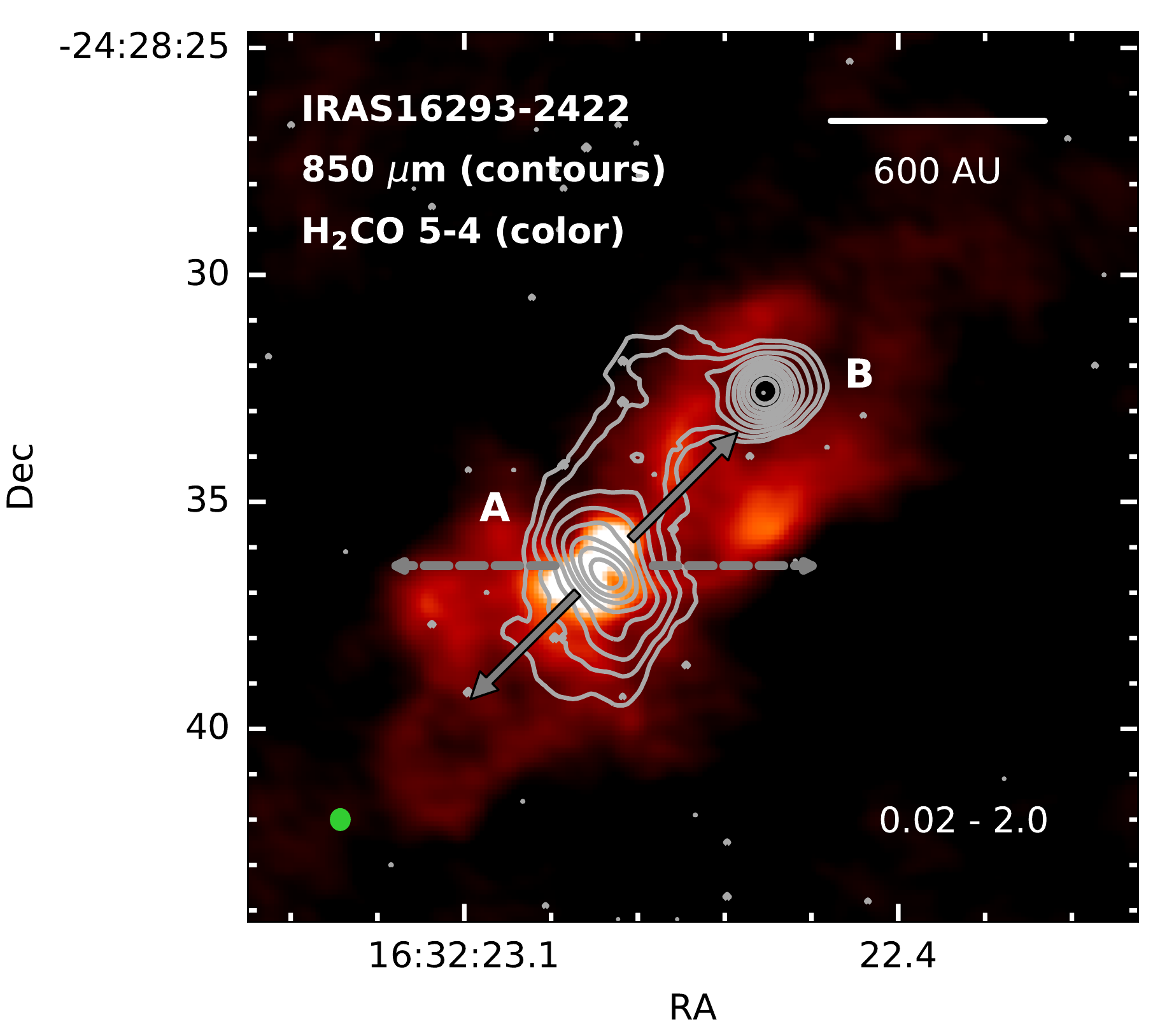}
	\includegraphics[width=0.5\textwidth]{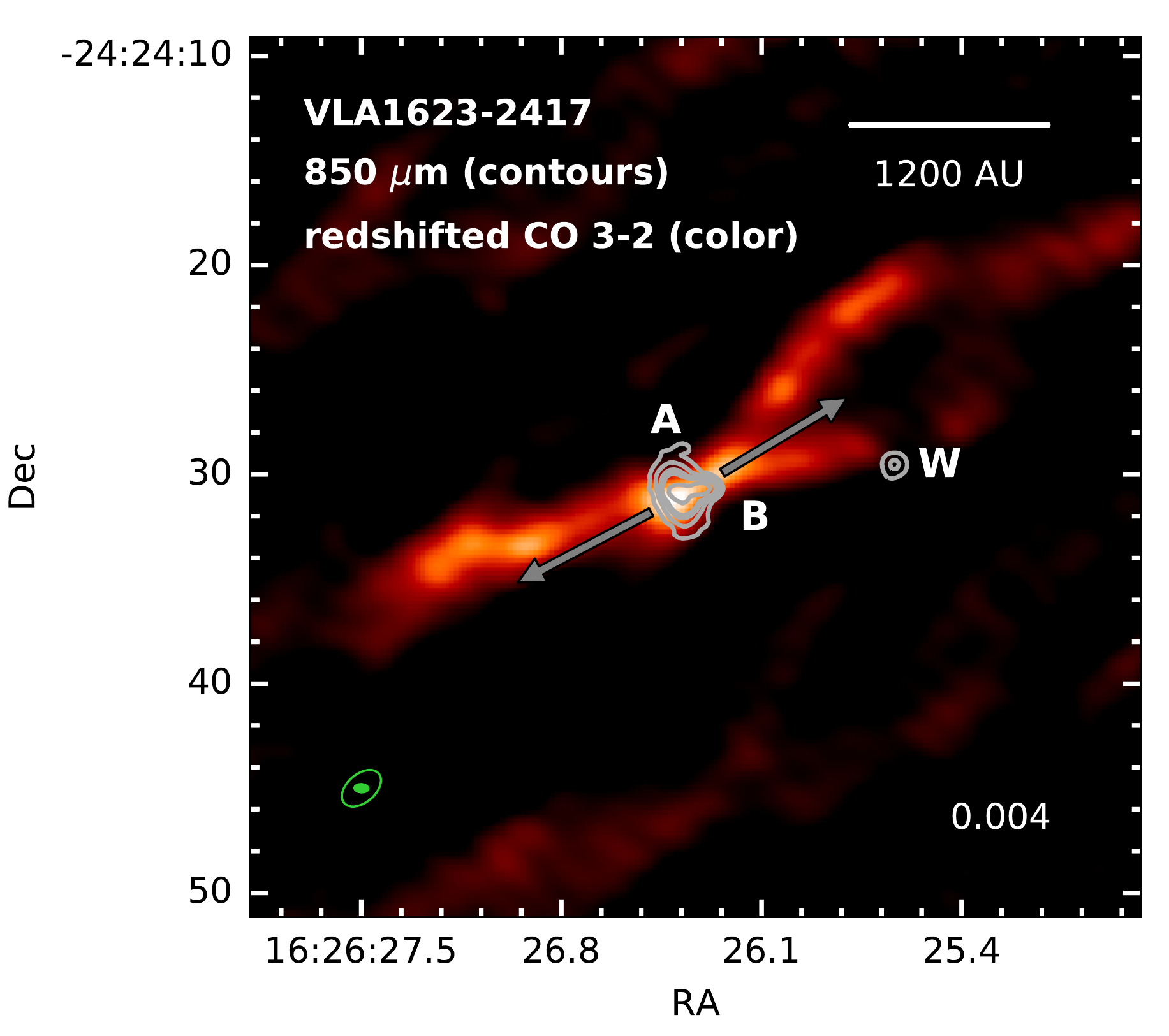}
	\caption{Continuum at 850$\mu$m (contours) for both systems, IRAS~16293$-$2422 (left) and VLA~1623$-$2417 (right), overlaid on intensity integrated \ce{H2CO} (0 to 60 km~s$^{-1}$; color-scale) and redshifted \ce{CO} (4 to 15 km~s$^{-1}$; color-scale), respectively. The green ellipses on the bottom left indicate the beam of the observations. For the right panel, the empty ellipse is the beam of the \ce{CO} observations. For VLA~1623$-$2417, contours are in steps of 3, 8, 15, 20 and 50$\sigma$, with $\sigma$ = 0.004~Jy~beam$^{-1}$. For IRAS~16293$-$2422, the levels are logarithmically spaced between 0.02 and 2~Jy~beam$^{-1}$, and highlight the ridge that spans between sources A and B. The arrows show the direction of red- and blue-shifted outflows from source A in each system.} 
	\label{fig:contsources}
\end{figure*}

\begin{figure*}
	\centering
	\includegraphics[width=\textwidth]{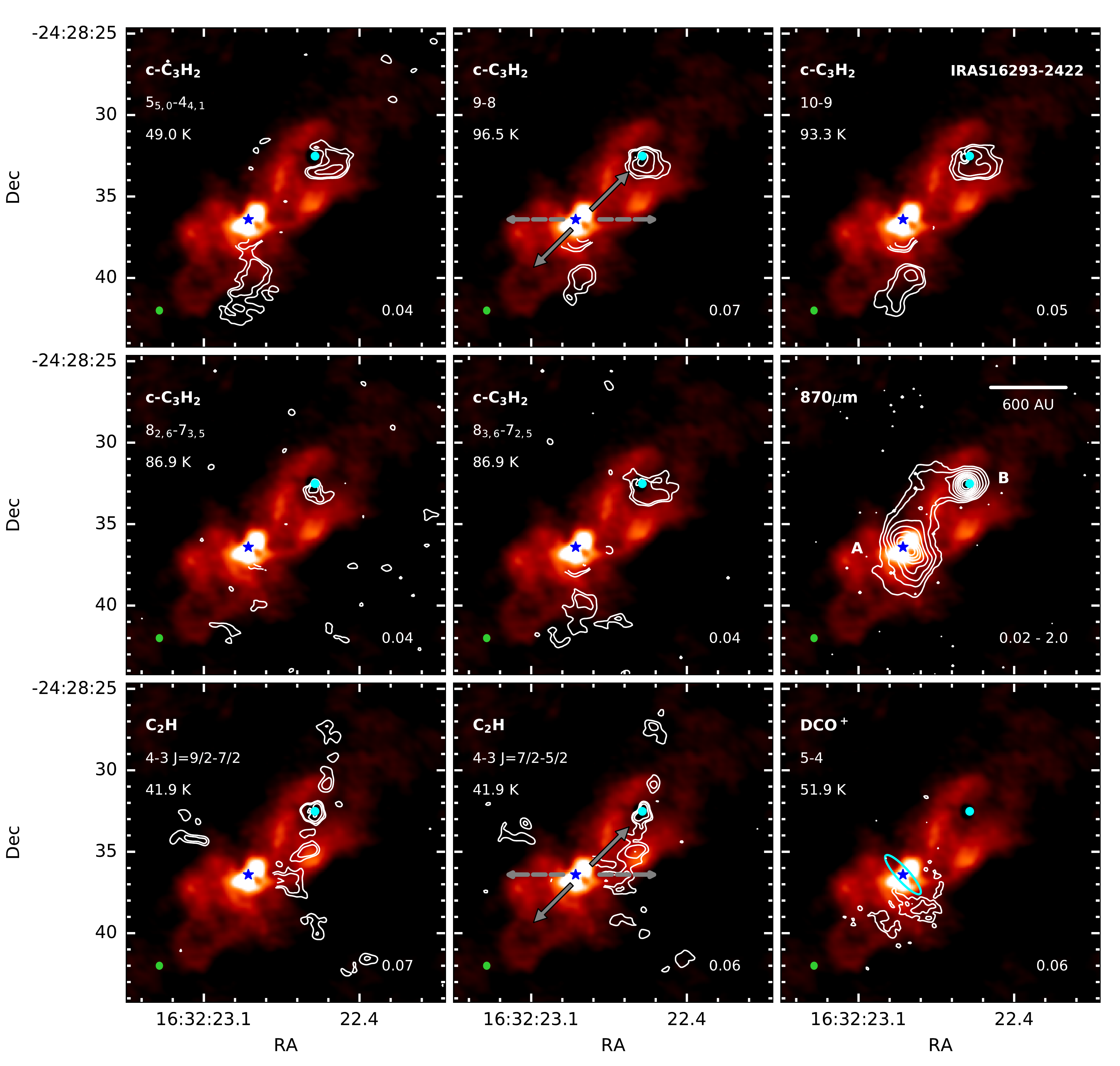}
	\caption{Intensity integrated maps (contours) of \ce{c-C3H2}, \ce{C2H}, \ce{DCO+} and continuum towards IRAS~16293-2422. Intensity integrated \ce{H2CO} 5--4 (0 to 60 km~s$^{-1}$) is shown in color-scale. Contours show the respective lines in steps of 2, 3, 5, 20 and 60$\sigma$, with $\sigma$ (Jy~beam$^{-1}$~km~s$^{-1}$) indicated in the lower right of each panel. For continuum, the levels are logarithmically spaced between 0.02 and 2~Jy~beam$^{-1}$, and highlight the ridge that spans between sources A and B. The positions of IRAS~16293-2422 A and B are indicated with a star and circle, respectively. The gray arrows indicate the outflow directions, while the cyan ellipse shows the disk-like structure. The green circle on the bottom left indicates the beam of the combined 12m and ACA observations. For the \ce{c-C3H2} and \ce{C2H} panels, the emission centered on A is contamination from other molecule(s) and is masked out in a radius of 2$\arcsec$ from the position of A.}
	\label{fig:i16293mom}
\end{figure*}

\begin{figure}
	\centering
	\includegraphics[width=\columnwidth]{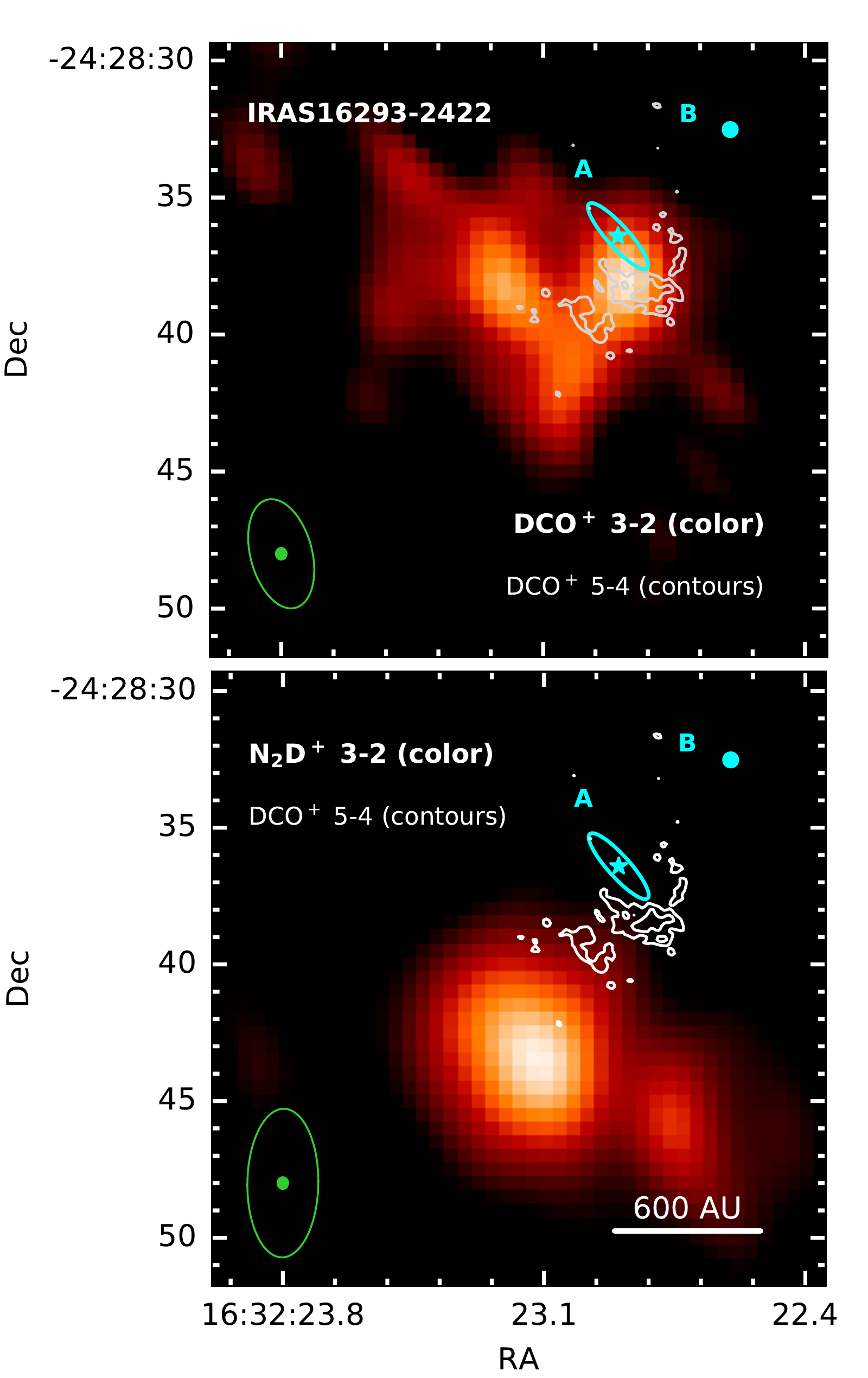}
	\caption{Intensity integrated maps of \ce{DCO+} 3--2 (top) and \ce{N2D+} 3--2 (bottom) observed with the SMA, overlaid with \ce{DCO+} 5--4 (contours) observed with ALMA (combined 12m array and ACA). Contours are the same as in Fig.~\ref{fig:i16293mom} with $\sigma$ = 0.06~Jy~beam$^{-1}$. The positions of IRAS~16293-2422 A and B are indicated with a star and circle, respectively. the cyan ellipse shows the disk-like structure. The filled green ellipses show the beam for the ALMA observations, while the unfilled ellipse shows the beam for the SMA observations. Both \ce{DCO+} transitions match spatially, and \ce{N2D+} is located beyond the extent of the \ce{DCO+} emission. Note the different center of this figure compared to Fig.~\ref{fig:i16293mom}.}
	\label{fig:i16293SMA}
\end{figure}

\begin{figure*}
	\centering
	\includegraphics[width=0.8\textwidth]{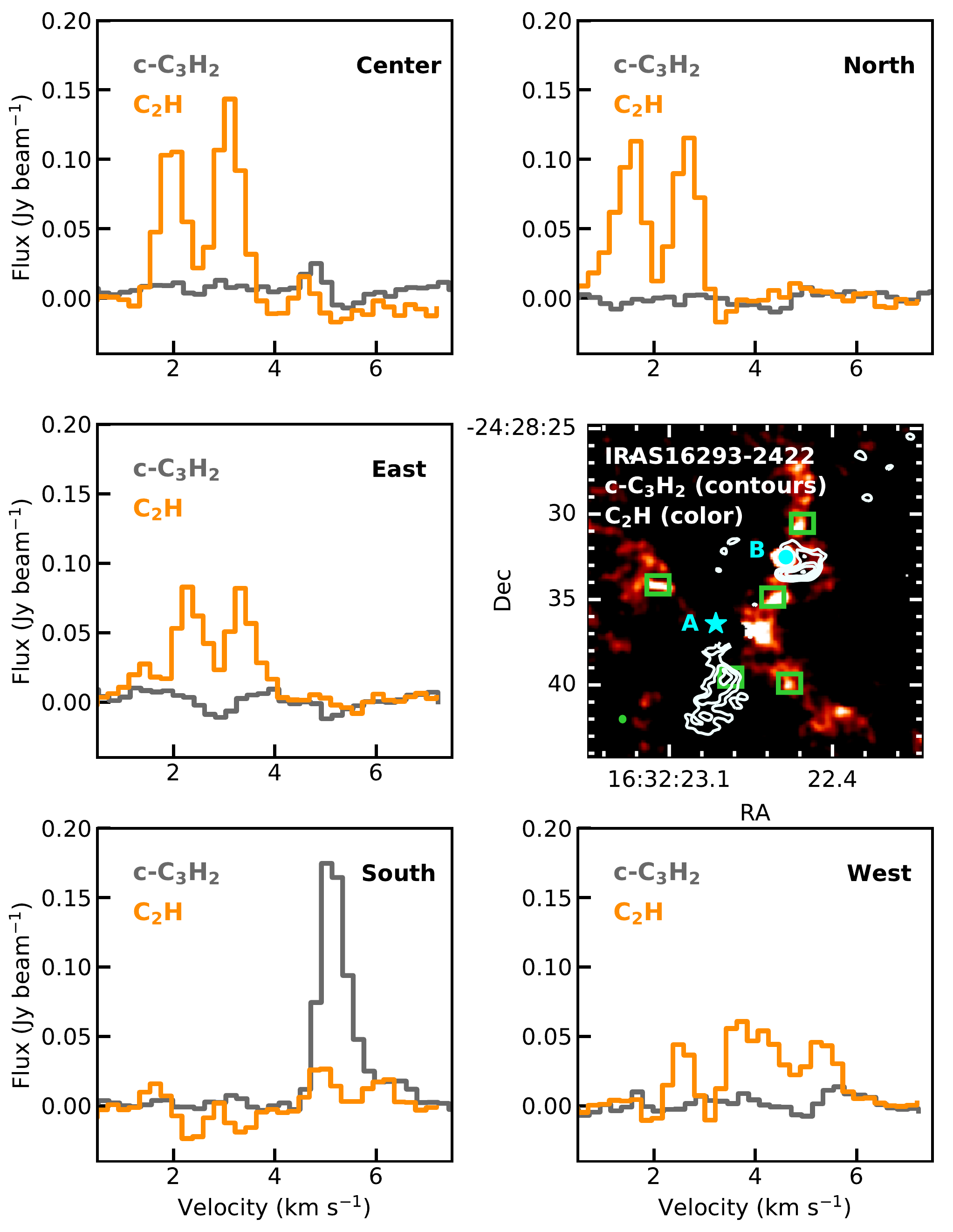} %
	\caption{IRAS~16293-2422 \ce{c-C3H2} 5--4 and \ce{C2H} 4--3 J=9/2--7/2 spectra taken at 5 positions, indicated by the green boxes in the intensity integrated map presented in the center right panel, while the green circle on the bottom left indicates the beam of the combined 12m and ACA observations. The anti-correlation of both molecules is seen at all positions. The region within a radius of 2$\arcsec$ from the position of source A is contaminated by other molecular species, and is masked out for these maps.}
	\label{fig:i16293mapspec}
\end{figure*}

\begin{figure*}
	\centering
	\includegraphics[width=\textwidth]{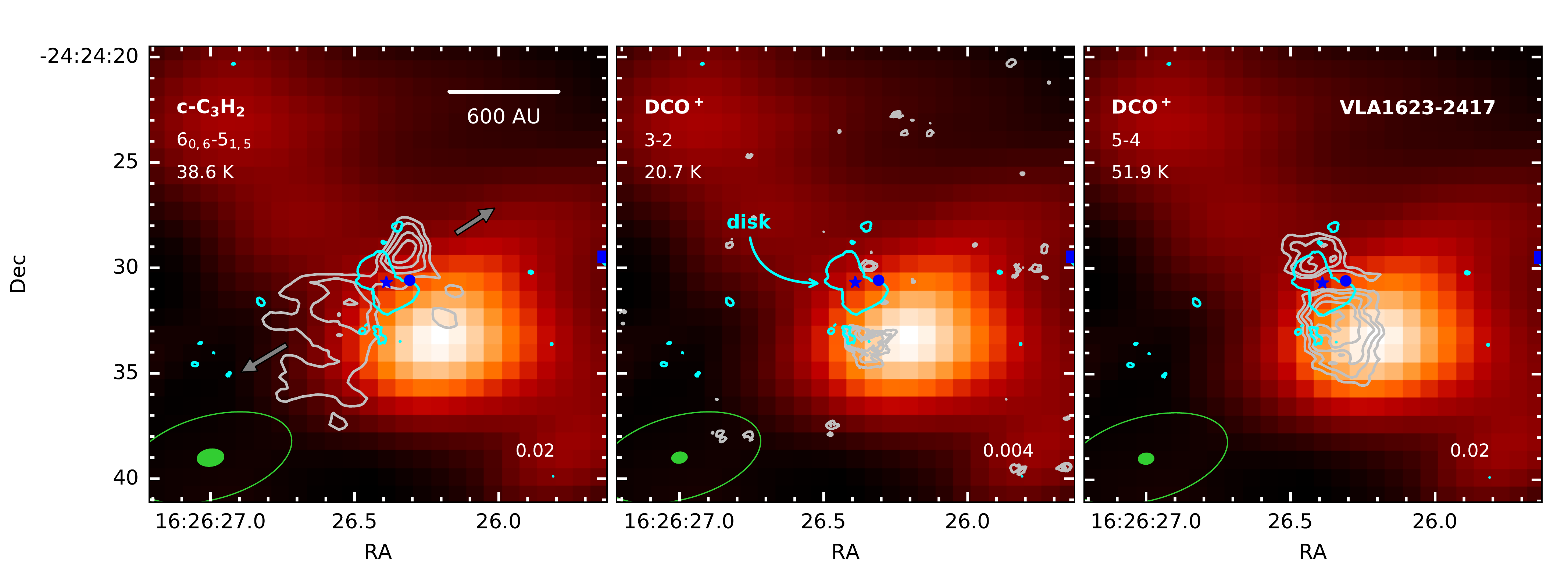}
	\caption{Intensity integrated maps of (white contours) of \ce{c-C3H2} and \ce{DCO+} (12m array) towards VLA~1623-2417. \ce{DCO+} 3--2 from ACA observations is shown in color-scale. The gray arrows on the left panel show the outflow direction. The cyan line is the 3$\sigma$ contour of \ce{C^{18}O} in order to show the location and extent of the rotating disk centered on VLA~1623-2417 A. Gray contours show the respective lines in steps of 3, 4, 5 and 6$\sigma$, except for \ce{DCO+} 3--2 where the contours start at 4$\sigma$. The value of $\sigma$ (Jy~beam$^{-1}$~km~s$^{-1}$) is indicated in the lower right of each panel. The positions of VLA~1623-2417 A, B and W are indicated with a star, circle and square, respectively. The filled green ellipses show the beam for the 12m array observations (contours), while the unfilled ellipse shows the beam of the ACA observation (color-scale).}
	\label{fig:vla1623mom}
\end{figure*}

\begin{figure}
	\centering
	\includegraphics[width=\columnwidth]{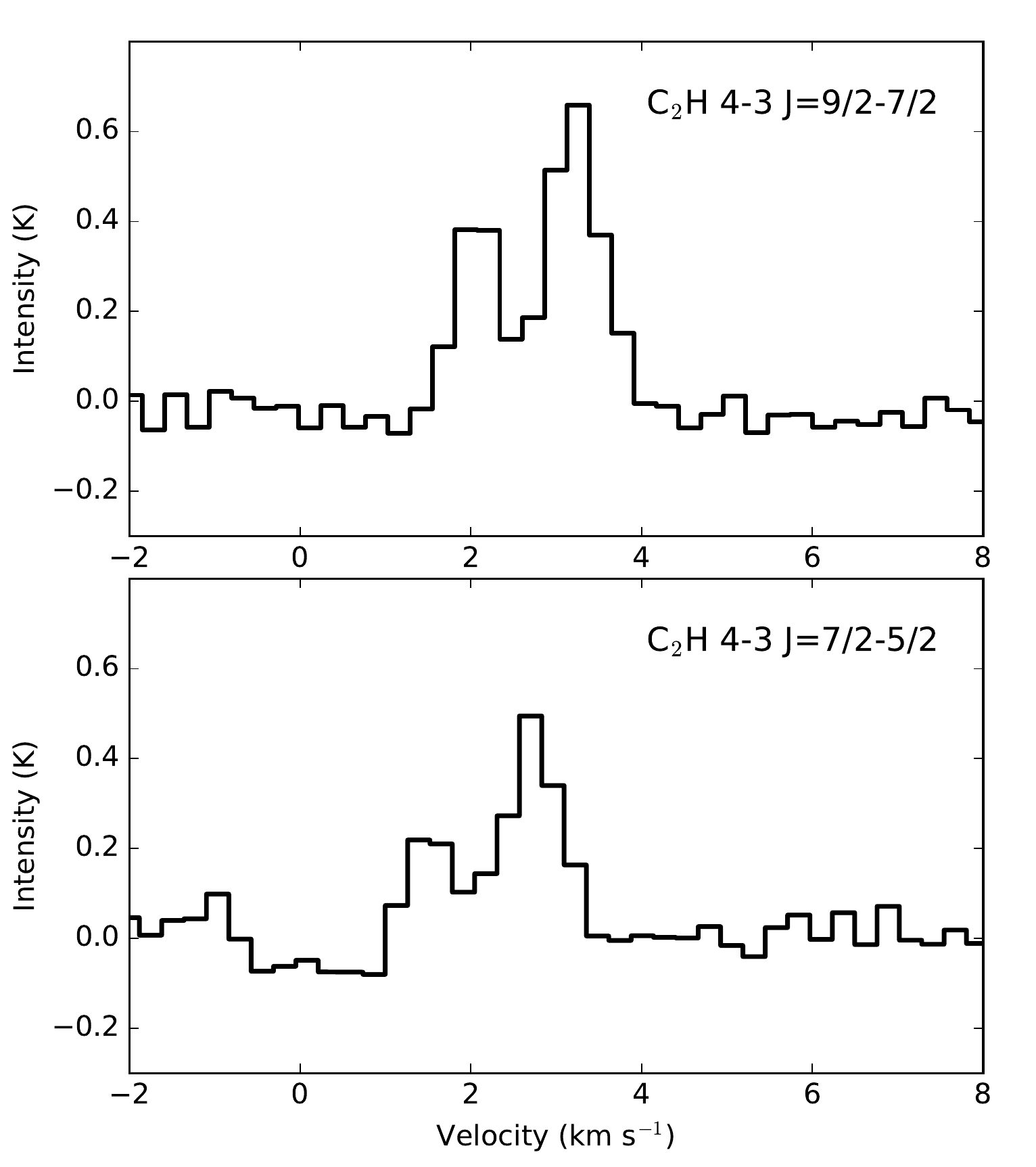}
	\caption{Single pointing APEX observations of \ce{C2H} centered on VLA~1623 A.}
	\label{fig:vla1623APEX}
\end{figure}

\begin{figure}
	\includegraphics[width=\columnwidth]{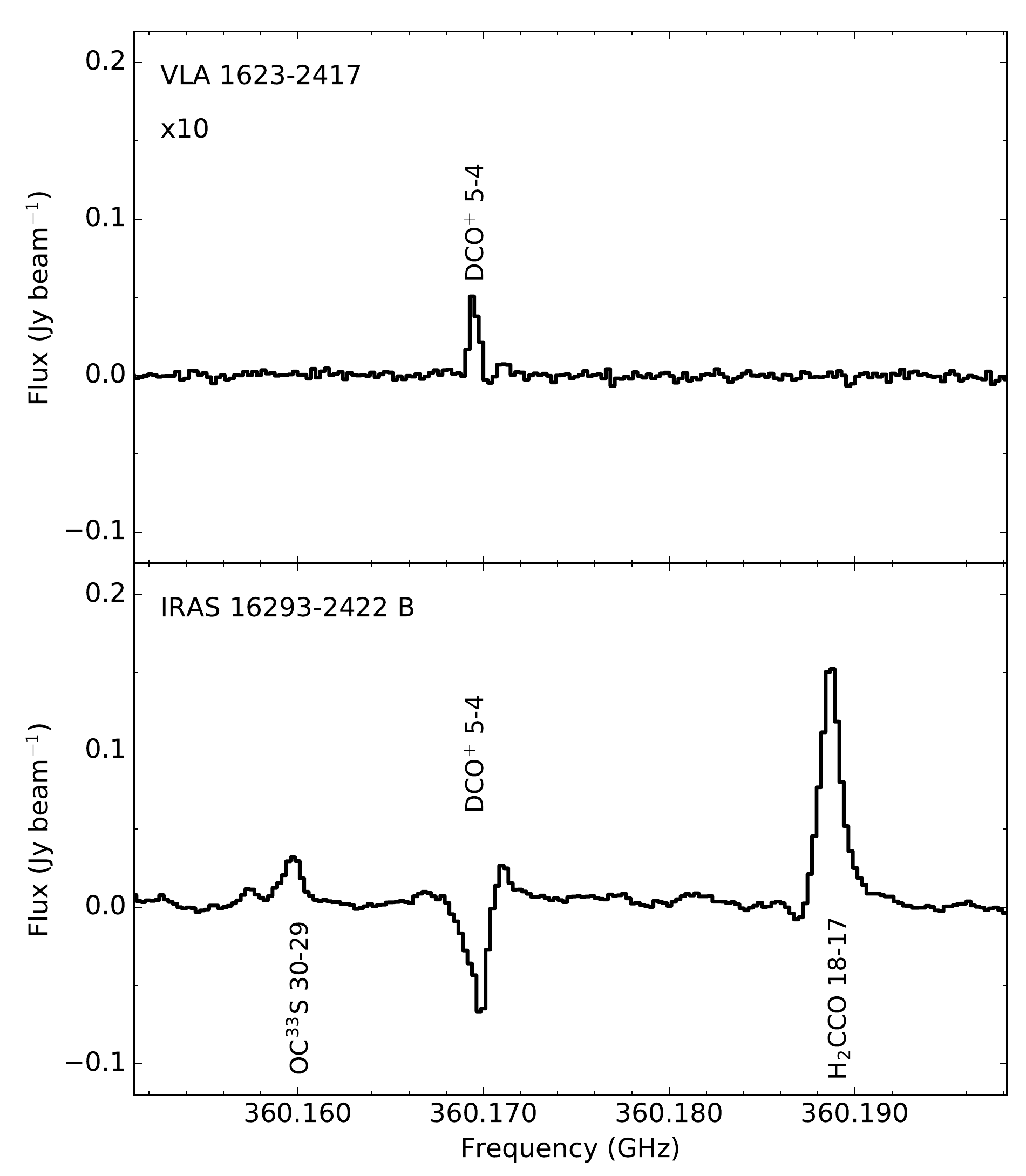}
	\caption{Comparison of the spectra centered on \ce{DCO+} 5--4 of VLA~1623-2417 A (12m array) and one beam offset from IRAS~16293-2422 B (combined 12m array and ACA). The spectra for VLA~1623-2417 has been multiplied by a factor of 10 in order to compare more easily. Note the lack of molecular line emission toward VLA~1623-2417 in contrast to IRAS~16293-2422.}
	\label{fig:compspec}
\end{figure}

\begin{table*} 
	\centering 
	\caption{Summary of line observations.}
	\label{tab:lines}
	\begin{tabular}{c c c c c c c c c c c}
		\hline \hline 
		& & & & & & \multicolumn{2}{c}{IRAS~16293-2422} & & \multicolumn{2}{c}{VLA~1623-2417} \\
		\cline{1-5}
		\cline{7-8}
		\cline{10-11}
		Line & Transition & $\nu$ & log$_{10}$ A$_{\rm ij}$ & E$_{\rm up}$ & & Peak Intensity & Line width & & Peak Intensity & Line width \\
		& & GHz &  & K & & mJy~beam$^{-1}$ & km~s$^{-1}$ & & mJy~beam$^{-1}$ & km~s$^{-1}$ \\
		\hline  
		\multicolumn{11}{c}{ALMA} \\
		\hline
		\ce{c-C3H2} & 6$_{0,6}$--5$_{1,5}$ & 217.822148 & -3.23 & 38.61 & & ... & ... && 194 & 0.5 \\
		\ce{c-C3H2} & 5$_{5,0}$--4$_{4,1}$ & 349.26400 & -2.78 & 48.98 && 240\tablefootmark{a} & 0.5 && ... & ... \\
		\ce{c-C3H2} & 10--9\tablefootmark{b} & 351.78158 & -2.61 & 96.49 && 410\tablefootmark{a} & 0.6 && ... & ... \\
		\ce{c-C3H2} & 9--8\tablefootmark{c} & 351.96597 & -2.67 & 93.34 && 350\tablefootmark{a} & 0.6 && ... & ... \\
		\ce{c-C3H2} & 8$_{2,6}$--7$_{3,5}$ & 352.18554 & -2.76 & 86.93 && 90\tablefootmark{a} & 0.6 && ... & ... \\
		\ce{c-C3H2} & 8$_{3,6}$--7$_{2,5}$ & 352.19364 & -2.76 & 86.93 && 200\tablefootmark{a} & 0.6 && ... & ... \\
		\ce{C2H} & 4--3 J=9/2--7/2 F=5--4 & 349.33771 & -3.88 & 41.91 && 170 & 0.6 && ... & ... \\
		\ce{C2H} & 4--3 J=9/2--7/2 F=4--3 & 349.33899 & -3.89 & 41.91 && 140 & 0.6 && ... & ... \\
		\ce{C2H} & 4--3 J=7/2--5/2 F=4--3 & 349.39927 & -3.90 & 41.93 && 140 & 0.6 && ... & ... \\
		\ce{C2H} & 4--3 J=7/2--5/2 F=3--2 & 349.40067 & -3.92 & 41.93 && 110 & 0.6 && ... & ... \\
		\ce{DCO+} & 3--2 & 216.11258 & -2.62 & 20.74 && ... & ... && 90 & 0.7 \\
		\ce{DCO+} & 5--4 & 360.16978 & -2.42 & 51.86 && 10 & 1.0 && 290 & 0.7 \\
		\ce{N2D+} & 3--2 & 231.32166 & -2.66 & 22.20 && ... & ... && $<$8.58\tablefootmark{d} & ... \\
		\ce{N2H+} & 4--3 & 372.67251 & -2.51 & 44.71 && ... & ... && $<$94.9\tablefootmark{d} & ... \\
		\hline  
		\multicolumn{11}{c}{SMA} \\
		\hline
		\ce{DCO+} & 3--2 & 216.11258 & -2.62 & 20.74 && 1800 & 1.0 && ... & ... \\
		\ce{N2D+} & 3--2 & 231.32166 & -2.66 & 22.20 && 1700 & 2.0 && ... & ... \\
		\hline  
		\multicolumn{11}{c}{APEX (T$_{\rm mb}$)} \\
		\hline
		\ce{C2H} & 4--3 J=9/2--7/2 F=5--4 & 349.33771 & -3.88 & 41.91 && ... & ... && 0.96 K & 0.7 \\
		\ce{C2H} & 4--3 J=9/2--7/2 F=4--3 & 349.33899 & -3.89 & 41.91 && ... & ... && 0.62 K & 0.7 \\
		\ce{C2H} & 4--3 J=7/2--5/2 F=4--3 & 349.39927 & -3.90 & 41.93 && ... & ... && 0.68 K & 0.7 \\
		\ce{C2H} & 4--3 J=7/2--5/2 F=3--2 & 349.40067 & -3.92 & 41.93 && ... & ... && 0.34 K & 0.7 \\
		\ce{DCO+} & 3--2 & 216.11258 & -2.62 & 20.74 && ... & ... && 4.8 K & 0.8 \\
		\ce{DCO+} & 5--4 & 360.16978 & -2.42 & 51.86 && ... & ... && 2.2 K & 0.8 \\
		\hline
	\end{tabular}
	\\
	\tablefoot{
		\tablefoottext{a}{\ce{c-C3H2} peak intensities and line widths taken from the south peak where there is no line confusion.}
		\tablefoottext{b}{Blended 10$_{0,10}$--9$_{1,9}$ and 10$_{1,10}$--9$_{0,9}$ transitions of \ce{c-C3H2}.}
		\tablefoottext{c}{Blended 9$_{1,8}$--8$_{2,7}$ and 9$_{2,8}$--8$_{1,7}$ transitions of \ce{c-C3H2}.}
		\tablefoottext{d}{1 $\sigma$ noise level of \ce{N2D+} and \ce{N2H+} in 0.02 km~s$^{-1}$ channel}}
	\tablebib{All rest frequencies were taken from the Cologne Database for Molecular Spectroscopy (CDMS) \citep{CDMS_2016}. The \ce{c-C3H2} 
		entry was based on \citet{c-C3H2_isos_rot_1987} with transition frequencies 
		important for our survey from \citet{c-C3H2_rot_1986} and from \citet{c-C3H2_isos_rot_2012}. 
		The CCH entry is based on \citet{CCH_obs_par_2009} with additional important data 
		from \citet{CCH_rot_2000} and \citet{CCH_rot1981}. 
		The \ce{DCO+} and \ce{N2H+} entries are based on \citet{DCO+_rot_2005} 
		and on \citet{cations_rot_2012}, respectively. Information on the \ce{N2D+} rest frequency was taken from 
		\citet{N2H+_N2H+_obs_freqs_2009}.}
\end{table*}

\section{Observations}
\label{sec:obs}
\subsection{IRAS~16293-2422}
IRAS~16293 was targeted in the ``Protostellar Interferometric Line Survey''
(PILS) program (Project-ID: 2013.1.00278.S; PI: Jes K. J{\o}rgensen;
\citealt{jorgensen2016}), an ALMA Cycle 2 unbiased spectral survey in
Band 7, using both the 12m array and the Atacama Compact Array (ACA).
The spectral set-up covers a frequency range from 329.147 GHz to
362.896 GHz, and provides a velocity resolution of 0.2
km~s$^{-1}$. The phase center was $\alpha_{J2000}$~=~16:32:22.72; $\delta_{J2000}$~=~$-$24:28:34.3, set to be equidistant
from the two sources A and B at $v_{\rm lsr}$ = 3.1 and 2.7 km~s$^{-1}$ \citep{jorgensen2011},
respectively. 
The resulting $(u,v)$ coverage of the combined 12m array and the ACA observations are sensitive to the distribution of material with an extent of up to 13$\arcsec$ and a circular synthesized beam of 0.5$\arcsec$.
A detailed description of the
observations and reduction is given in \cite{jorgensen2016}. 
Many complex molecules detected in this survey (e.g., \citealt{coutens2016,jorgensen2016,lykke2017})
show relatively compact emission peaking close to the location of
the two protostars. This
work focuses on a few of the molecules detected in the spectral survey
that show extended emission, namely \ce{DCO+}, \ce{c-C3H2} and
\ce{C2H}. 
Several transitions of \ce{c-C3H2} are present throughout the frequency range of PILS, however for this work, only a five transitions with strong observed emission were chosen in the 349 to 352 GHz frequency range.
For all molecules the combined 12m array and ACA
data cubes are used.  The typical RMS noise is about
7--10 mJy~beam$^{-1}$ per 0.2 km s$^{-1}$ channel, and the flux
calibration uncertainty is $\sim$5\% \citep{jorgensen2016}.  Transitions and
line frequencies of the molecular species used in this work are listed
in Table~\ref{tab:lines}, as well as the peak intensities and line widths.
Table~\ref{tab:uv} lists further details of the observations, such as 
UV-baseline range and largest angular scale recovered in the observations.

Since the PILS survey did not cover lines of \ce{N2D+} 3--2 or \ce{DCO+}
3--2, we include here the observations of these two molecules from a
spectral line survey with the SMA \citep{jorgensen2011}.
The phase center was $\alpha_{J2000}$~=~16:32:22.91; $\delta_{J2000}$~=~$-$24:28:35.5.
For \ce{DCO+} 3--2, the beam size is 5.5$\arcsec \times$ 3.2$\arcsec$ (P.A. = 17.7$^{\circ}$), while for \ce{N2D+} 3--2 the beam size is 4.0$\arcsec \times$ 2.4$\arcsec$ (P.A. = $-$1.0$^{\circ}$). 
The RMS noise is 0.24~Jy~beam$^{-1}$ for a 0.56 km s$^{-1}$ channel width for \ce{DCO+}, and 0.06~Jy~beam$^{-1}$ for a channel width of 1.1 km s$^{-1}$ for \ce{N2D+}.
Further details on the reduction and analysis are given in \citet{jorgensen2011}.
These data are considered in order to directly compare the cold chemistry of IRAS~16293
with that of VLA~1623.  Additionally, two transitions of \ce{DCO+} are needed
to derive temperature and density from line ratios.

\subsection{VLA~1623-2417}
VLA~1623 was observed with ALMA in Cycle 0 using Band 6, with phase center $\alpha_{J2000}$~=~16:26:26.419; $\delta_{J2000}$~=~$-$24:24:29.988.
The spectral set-up was configured to observe \ce{DCO+} 3--2 and \ce{N2D+} 3--2 together with \ce{C^{18}O} 2--1 and \ce{^{12}CO} 2--1, providing a velocity resolution of 0.0847 km~s$^{-1}$ and a synthesized beam size of 0.85$\arcsec$$\times$0.56$\arcsec$ (P.A. = $-$83.8$^{\circ}$).
\ce{DCO+} 3--2 data from the Cycle 0 observations were previously presented in \cite{murillo2015} and are added to this work for completeness in the comparison of both systems.
The data reduction results of \ce{C^{18}O} and \ce{^{12}CO} can be found in \cite{murillo2013} and \cite{santangelo2015}.

ALMA Cycle 2 observations of VLA 1623 were carried out in Band 6 and 7, with phase center $\alpha_{J2000}$~=~16:26:26.390; $\delta_{J2000}$~=~$-$24:24:30.688. Baseline and frequency ranges are listed in Table~\ref{tab:uv}.

The Cycle 2 Band 6 spectral set-up covered \ce{DCO+} 3--2, \ce{C^{18}O} 3--2, \ce{^{13}CO} 3--2 and \ce{c-C3H2} 6$_{0,6}$--5$_{1,5}$ together with continuum.
Data calibration was done with J1733-1304 and J1517-2422 for bandpass, J1625-2527 for gain calibration, and J1517-243, J1733-130, Ceres, Mars, and Titan were observed for flux calibration.
The spectral windows with line emission have bandwidths of 62 MHz each, while for continuum the total bandwidth is of 8 GHz.
These observations were carried out with both the Total power, ACA and 12m arrays, for a total of four configurations.
With the 12m array, two configurations, C35-5 (25 $\sim$ 1000 m) and C34-1 (10 $\sim$ 350 m), were used in order to bridge the gap between the 12m and ACA array observations.
Cycle 2 Band 6 ACA observations of \ce{DCO+} were presented in \cite{murillo2015} and are only included here for visual comparison.
The 12m and ACA \ce{C^{18}O} and \ce{^{13}CO} observations from Cycle 2 Band 6 will be treated in a future publication.
In this work, we focus only on the 12m array \ce{c-C3H2} 6$_{0,6}$--5$_{1,5}$ observations from Cycle 2 Band 6 data.
The C35-5 configuration provides an angular and velocity resolution of 0.45$\arcsec$$\times$0.25$\arcsec$ (P.A. = 86.3$^{\circ}$) and 0.0208 km~s$^{-1}$, respectively, with a typical RMS noise of 7 mJy beam$^{-1}$.
The C34-1 configuration results in an angular resolution of 1.60$\arcsec$$\times$0.88$\arcsec$ (P.A. = 83.8$^{\circ}$) with a channel width of 0.0208 km~s$^{-1}$ and a typical RMS noise of 20 mJy beam$^{-1}$.

Band 7 observations, with a spectral set-up covering \ce{N2H+} 5--4,
\ce{DCO+} 5--4 and \ce{H2D+} 1$_{1,0}$--1$_{1,1}$ as well as
continuum with only the 12m array, provided a spectral and angular
resolution of 0.025 km~s$^{-1}$ and 0.88$\arcsec$$\times$0.56$\arcsec$ (P.A. = $-$86.6$^{\circ}$),
respectively. The spectral windows with line emission have a total bandwidth of 62 MHz each, and continuum has a total bandwidth of 4 GHz. Total observing time was 0.9 hr with a 46\% duty
cycle, using 34 antennas and a maximum baseline of 350 m.  Data
calibration was done with J1517-2422, J1625-2527 and Titan for
bandpass, gain and flux calibration, respectively. \ce{DCO+} was detected
with a noise of 26 mJy~beam$^{-1}$ per 0.025 km~s$^{-1}$. The system
temperature was relatively high for the spectral windows containing
\ce{N2H+} and \ce{H2D+}, causing the noise to be of about 95
mJy~beam$^{-1}$ per 0.025 km~s$^{-1}$ velocity channel, despite flagging
the antennas with the highest system temperature.

In this work we focus on the \ce{DCO+} 3--2 and
\ce{N2D+} 3--2 lines from the Cycle 0 12m array obsevations, in addition to
\ce{c-C3H2} 6$_{0,6}$--5$_{1,5}$, \ce{DCO+} 5--4, \ce{N2H+} 5--4 and \ce{H2D+} 1$_{1,0}$--1$_{1,1}$
12m array observations from Cycle 2.  Line transitions and frequencies together
with peak intensities and line widths are listed in Table~\ref{tab:lines}.

Additionally, single-dish APEX observations in the ON/OFF mode were
carried out on 22 and 24 October 2016 using the heterodyne instrument
SheFI \citep{belitsky2006,vassilev2008} with Bands APEX-1 (213 -- 275 GHz) 
and APEX-2 (267 -- 378 GHz),
targeting \ce{DCO+} 3--2 and 5--4, as well as \ce{C2H} 4--3.  
These observations were taken to compare the location of \ce{C2H} in
both VLA~1623 and IRAS~16293, as well as to have a separate verification and
comparison of the physical parameters derived from ALMA observations
and single-dish.  
Several transitions of \ce{NO} and \ce{HCN} were detected, both of which
can form in gas and surface reactions, whereas \ce{N2H+} and \ce{N2D+} 
only form in the gas. \ce{NO} and \ce{HCN} are not further analyzed 
in this work. The observations were centered
on VLA~1623 A ($\alpha_{J2000}$ = 16:26:26.390; $\delta_{J2000}$ =
$-$24:24:30.688).  The typical RMS noise was 100 mK for APEX-1 and between
50 -- 80 mK for APEX-2 in 0.1 km~s$^{-1}$ channels.
Peak temperatures ($T_{\rm mb}$) and line widths for Gaussian fits to the single dish lines are listed in Table~\ref{tab:lines}. 
The typical calibration uncertainties are about 10\% for the APEX SheFI 
instruments in the 230 and 345 GHz Bands.  
For APEX-1 and APEX-2 observations the HPBW is 28.7$\arcsec$ and 18$\arcsec$, respectively.
The main beam efficiencies
used are $\eta_{\rm mb}$ = 0.75 at 230 GHz, and $\eta_{\rm mb}$ = 0.73
at 345 GHz. 

\begin{table*}
	\caption{Tests of the temperature and density profiles of the two sources.}
	\label{tab:tests}
	\centering 
	\begin{tabular}{c c c c c c c c}
		\hline \hline
		& \multicolumn{3}{c}{IRAS~16293-2422 \citep{crimier2010}} & & \multicolumn{3}{c}{VLA~1623-2417 \citep{jorgensen2002}} \\
		\cline{2-4}
		\cline{6-8}
		Test & $T_{\rm 27AU}$ (K) & $n_{\rm 27AU}$ (cm$^{-3}$) & Note && $T_{\rm 4AU}$ (K) & $n_{\rm 4AU}$ (cm$^{-3}$) & Note \\
		\hline
		1 & 300.0 & 2.36 $\times$ 10$^{9}$ & unchanged && 250.0 & 1.62 $\times$ 10$^{9}$ & unchanged \\
		2 & 300.0 & 2.36 $\times$ 10$^{10}$ & $n$ increased by 10 && 250.0 & 1.62 $\times$ 10$^{10}$ & $n$ increased by 10 \\
		3 & 300.0 & 2.36 $\times$ 10$^{8}$ & $n$ decreased by 10 && 250.0 & 1.62 $\times$ 10$^{8}$ & $n$ decreased by 10 \\
		4 & 100.0 & 2.36 $\times$ 10$^{9}$ & $T$ decreased by 3 && 166.7 & 1.62 $\times$ 10$^{9}$ & $T$ decreased by 1.5 \\
		\hline
	\end{tabular}
\end{table*}

\begin{table}
	\centering
	\caption{\ce{DCO+} best approximation model parameters.}
	\label{tab:dcomod}
	\begin{tabular}{c c c}
		\hline \hline
		Parameter & IRAS~16293-2422 & VLA~1623-2417 \\
		\hline
		\multicolumn{1}{c}{$T_{\rm peak}$ (K)} & 17--19 & 11--16 \\
		\multicolumn{1}{c}{Drop boundaries:} & & \\
		$T_{\rm sub}$ (K) & 35 & 35 \\
		$n_{\rm de}$ (cm$^{-3}$) & $\leq$10$^{6}$ & 3$\times$10$^{6}$ \\
		\multicolumn{1}{c}{\ce{CO} abundance:} & & \\
		Inner $X_{\rm in}$ & 10$^{-5}$ & 10$^{-5}$ \\
		Drop $X_{\rm D}$ & 10$^{-6}$ & 10$^{-7}$ \\
		Outer $X_{0}$ & 10$^{-4}$ & 10$^{-4}$ \\
		\hline
	\end{tabular}
\end{table}

\section{Results}
\label{sec:res}

\subsection{IRAS~16293-2422}
The molecules \ce{c-C3H2}, \ce{C2H} and \ce{DCO+} from the PILS
spectral survey \citep{jorgensen2016} are considered here, together
with \ce{DCO+} and \ce{N2D+} from the SMA spectral survey
\citep{jorgensen2011}. 
The PILS survey images are obtained from the combined 12m array and ACA, which picks up the small and large scale emission from scales less than 13$\arcsec$.
The peak intensities and widths of each
line are listed in Table~\ref{tab:lines}. Intensity integrated maps
of each line overlaid on \ce{H2CO} are shown in
Fig.~\ref{fig:i16293mom}. \ce{H2CO} 5$_{1,5}$--4$_{1,4}$ from the
PILS survey (van der Wiel et al. in prep.) is used as a reference for the more
extended envelope and one of the outflow directions.  The nominal
velocities at which most species emit at source A and B are $V_{\rm
  LSR}=$3.2 and 2.7 km s$^{-1}$. \ce{c-C3H2} and \ce{C2H} spectra at selected
positions are presented in Fig.~\ref{fig:i16293mapspec} and
\ref{fig:i16293spec}.

\ce{DCO+} is detected in the 5--4 (PILS) and 3--2 (SMA) transitions,
with a half-crescent shape centered around source A (Fig.~\ref{fig:i16293mom} and ~\ref{fig:i16293SMA}).
The peak is red-shifted and located $\sim$2$\arcsec$ southwest of
source A in both transitions, consistent with the red-shifted emission of the disk-like structure to the southwest \citep{oya2016}.  
Weak absorption is detected towards source B in the PILS observations, which is consistent with previous studies that indicate infall motions through an inverse P-cygni profile \citep{zapata2013}. 
The \ce{DCO+} emission south of source A is weak, peaking at 3
$\sigma$ in the 5--4 transition and at 5$\sigma$ in the 3--2 transition.  
It is slightly extended to the
south along the outflow, but not as far as \ce{c-C3H2}.  
In agreement with previous 
observations of \ce{DCO+} and \ce{c-C3H2} for other objects \citep{spezzano2016a,spezzano2016b}, these two
molecules are spatially anti-correlated.

Five narrow (FWHM $\approx 1$ km s$^{-1}$) lines of \ce{c-C3H2} in the 349 to 352 GHz frequency range with
$E_{\rm up}$ ranging from 48 to 96 K are studied in this work.  The emission peaks to
the south of A, seen clearly in the top row of
Fig.~\ref{fig:i16293mom}. The lines are also seen
near source B, at one ALMA beam offset from the
source. 
The southern emission extends from the circumstellar region of source
A, and peaks at $\leq$5$\sigma$ about $\sim$4$\arcsec$ away from the
source position.
Toward source A itself, the spectrum is too confused to identify the separate molecular lines.
Thus the region is masked out
in the maps within 2$\arcsec$ from the source position. 
Comparing \ce{c-C3H2} with \ce{H2CO} (Fig.~\ref{fig:i16293mom}) 
suggests that it could arise from one side of the southern outflow cavity wall. 
The asymmetric heating of the outflow cavity could be due to the behavior of the outflow from IRAS~16293 A.
Observations of the outflow of IRAS~16293 A at large and small scales suggest that the outflow has shifted direction
from east-west (dashed arrow shown in Fig.~\ref{fig:contsources}) to southeast-northwest (solid arrow shown in Fig.~\ref{fig:contsources}).
The shift of outflow direction could have swept up material, thus causing the asymmetric morphology of the outflow cavity, and consequently
\ce{c-C3H2} to only be present on one side of the outflow cavity. 
The emission around source B may either be 
from the circumstellar region or the outflow cavity, but due to the 
orientation it is difficult to say. 

\ce{C2H} is clearly detected in both spin doubling transitions with
each transition showing a characteristic double hyperfine structure
pattern. 
The emission within 2$\arcsec$ of source A is masked out due to contamination
from other molecular species.
\ce{C2H} emission is located in a filament-like structure extending from 
north to south, passing through source B (Fig.~\ref{fig:i16293mom}).
A second, weaker structure formed by a string of clumps extends from
north-east to south-west, apparently passing through source A.
\ce{C2H} is diffuse and weak, peaking at $\lesssim$5$\sigma$
on all off-source positions on the map (Fig.~\ref{fig:i16293mapspec}).
The emission around source B is brighter, peaking at 10$\sigma$ in the intensity integrated map.
From the channel map, the emission appears to have a subtle velocity gradient from north to south at source B.
However, the \ce{C2H} emission does not match the structure and
extent of \ce{c-C3H2} (or \ce{H2CO}) in either transition (Fig.~\ref{fig:i16293mapspec} and \ref{fig:i16293spec}).
It should be noted that neither \ce{C2H} or \ce{c-C3H2} coincide with the dust ridge
seen in the continuum emission (Fig.~\ref{fig:i16293mom}; see also \citealt{jacobsen2017}).

\ce{N2D+} 3--2 is mostly resolved out in the PILS survey, but it is 
detected south-east of \ce{DCO+} with the SMA (Fig.~\ref{fig:i16293SMA}), located 7$\arcsec$
away from the continuum position of source A with a S/N = 7 \citep{jorgensen2011}. Similar to \ce{DCO+}, there is
no \ce{N2D+} emission towards B.
No transition of \ce{N2D+} was covered in the Band 7 observations.

\subsection{VLA~1623-2417}

Two transitions of \ce{DCO+} and one transition each of \ce{c-C3H2},
\ce{N2H+} and \ce{N2D+} were observed with ALMA in Bands 6 and 7. 
Additional APEX observations
detected two transitions of \ce{DCO+} and \ce{C2H}.  Intensities and
line widths are listed in Table~\ref{tab:lines}.  Fig.~\ref{fig:vla1623mom} 
shows the intensity integrated maps for the ALMA observations. All of these molecules trace material
associated with VLA~1623 A, but not the other two components of the system,
VLA~1623 B and W.

The \ce{DCO+} 3--2 ALMA 12m array and ACA observations have been separately
analysed in detail in \cite{murillo2015}.  The 3--2 ACA map shows a
smooth distribution peaking south-west of the source, with the
blue-shifted emission extending north-east, but no clear red-shifted
counterpart south-west.  Here we present additional ALMA Band 7 12m
array observations of \ce{DCO+} 5--4.  In both transitions of
\ce{DCO+} the red-shifted emission, located to the south of VLA~1623 A, is
clearly seen and is stronger than the blue-shifted
emission located to the north.  The \ce{DCO+} 5--4 emission is three
times stronger than the 3--2 emission with the 12m array, which makes
the blue-shifted emission clearly visible.  For both transitions the
velocity gradients are consistent.  The \ce{DCO+} 3--2 emission
borders the disk structure observed to be driven by VLA~1623 A and is
relatively compact.  Even more
interesting, however, is that \ce{DCO+} in the 5--4 transition extends
closer to the position of VLA~1623 A than in the 3--2 transition
(Fig.~\ref{fig:vla1623mom}).

\ce{DCO+} forms at temperatures below 20 K, where \ce{CO} freezes out.
The position of the \ce{DCO+} 3--2 peak along the disk plane was found
to be the product of disk-shadowing, which causes a temperature drop
at the edge of the disk, whereas along the outflow direction no such
effect was observed \citep{murillo2015}.  The APEX observations of
\ce{DCO+} in both transitions show a single peak at the systemic
velocity (3.7 -- 4 km~s$^{-1}$) and a peak intensity of 3.6 K for the
3--2 transition, the same as obtained from JCMT observations by
\cite{jorgensen2004}. The
beamsize for the APEX-1 and 2 Bands covers approximately the full
extent of the \ce{DCO+} emission seen in the ACA map.
For the \ce{DCO+} 3--2 transition, the ALMA observations recover about
28\% of the flux detected in the APEX observations (117.1~Jy~km~s$^{-1}$
with 24~Jy/K), while 20\% was recovered with the \ce{DCO+} 5--4 ALMA 
observations (APEX: 53.7~Jy km~s$^{-1}$ with 24~Jy/K).

One low-lying transition of \ce{c-C3H2} is detected with the short baselines of the 12m array, but
not the long baselines. The detection of \ce{c-C3H2} with only the the short baselines of the 12m array indicates that the emission is extended without a compact structure component. 
The \ce{c-C3H2} emission is oriented perpendicular
to the disk and seems to trace the cavity of the outflow driven by
VLA~1623 A out to 3$\arcsec$ from the source position.  There is no
detection of \ce{c-C3H2} emission in the disk traced by \ce{C^{18}O} or
at the disk-envelope interface, down to the noise level.  The material
along the outflow cavity exhibits signatures of rotation, most notable
in the south-east lobe, with a velocity range and gradient direction
similar to that of \ce{DCO+} and \ce{C^{18}O} \citep{murillo2013}.
However, treating the kinematics of the outflow cavity wall traced
by \ce{c-C3H2} is outside the scope of this paper, and will be presented in a separate paper.

\ce{C2H} is detected with APEX, with the hyperfine components of each
transition being clearly distinguished (Fig.~\ref{fig:vla1623APEX}).  Both transitions are located
at the systemic velocity of VLA~1623 A (3.7 -- 4.0 km~s$^{-1}$) and show no
broadening, indicating that the emission is most likely related to the
envelope material of VLA~1623 A.

\ce{N2H+} and \ce{N2D+} are not detected in our ALMA observations.
Possible reasons could be either due to the emission being very
extended and thus resolved out in the interferometric observations, or
the abundance of these molecules being too low to be detected.  This
is a surprising contrast to several other young embedded Class 0
sources which do show \ce{N2H+} and \ce{N2D+}
\citep{jorgensen2004,tobin2013}. The non-detections of \ce{N2H+} and \ce{N2D+}
are further analysed in Sect.~\ref{sec:analysis:nitrogen}.  \ce{H2D+}
is also not detected in our Cycle 2 Band 7 observations, this is
consistent with the JCMT observations reported by \cite{friesen2014}.
\ce{H2D+} is not further treated in this work.

\begin{figure*} 
	\centering
	\includegraphics[width=0.9\textwidth]{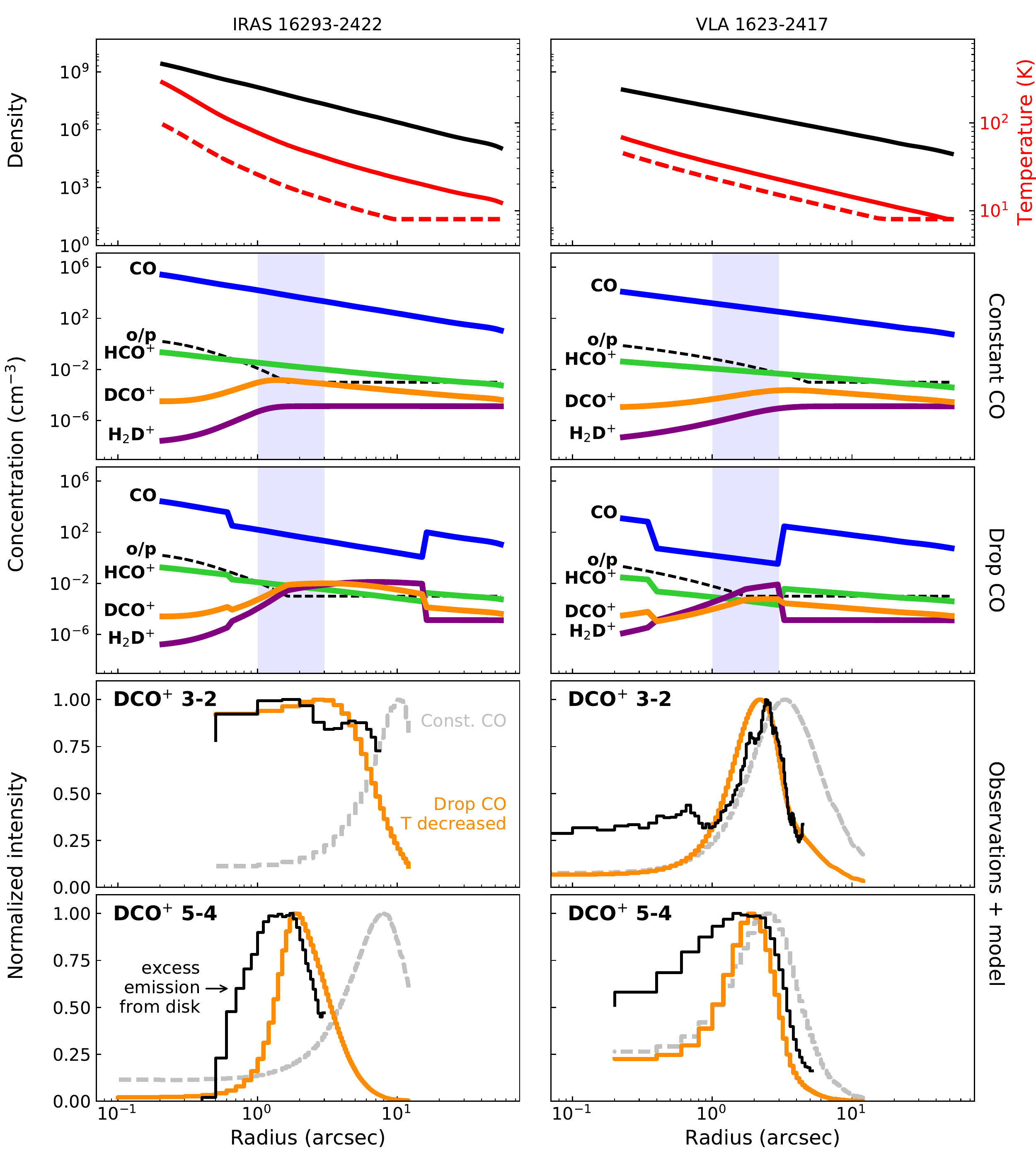}
	\caption{Results from modelling the observed \ce{DCO+} peak with our simple analytic chemical model. The left column shows results for IRAS~16293-2422, while the right column shows those for VLA~1623-2417. The top row shows the input density (black) and temperature (red) as functions of radius for each system. The solid red line shows the original temperature profile used for the constant CO model (second row panels). The dashed red line is the profile decreased by factors of 3 and 1.5 for IRAS~16293-2422 and VLA~1623-2417, respectively, and used for the drop CO model (third row panels). The lower limit on the temperature is set at 8K. In the second and third row, the shaded range is the location of the observed \ce{DCO+} peak. The fourth and fifth rows show the observed radial profiles of \ce{DCO+} (black lines) overlaid with both Constant (dashed gray lines) and Drop CO models (solid orange lines). In all panels, the protostellar source is located on the left, and the envelope on the right, with the peak position of \ce{DCO+} indicating the disk-envelope interface. Note that the excess \ce{DCO+} 5--4 emission from the disk(-like) regions is not well reproduced by the cold \ce{DCO+} network.}
	\label{fig:iras16293chemnet}
\end{figure*} 

\begin{figure*}
	\centering
	\includegraphics[width=0.95\textwidth]{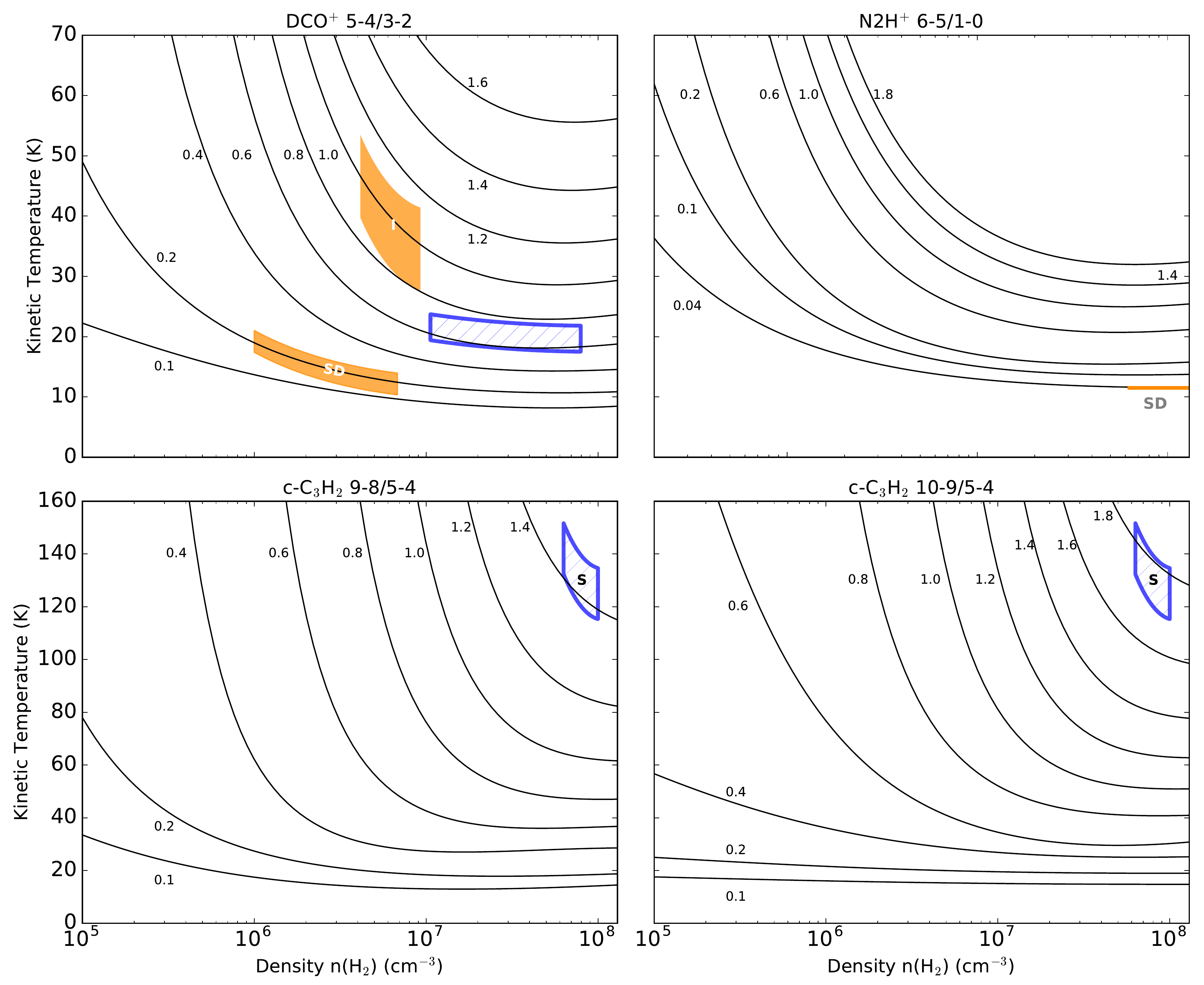}
	\caption{Calculated line brightness temperature ratios for \ce{DCO+} 5--4/3--2,
		\ce{N2H+} 6--5/1--0 and \ce{c-C3H2} 10--9/5--4 and
		9--8/5--4. Black lines show the modelled ratios assuming
		column densities of 2.5$\times$10$^{12}$ and
		1.3$\times$10$^{13}$ cm$^{-2}$ for \ce{DCO+} and \ce{N2H+},
		respectively, and 7$\times$10$^{13}$ for both \ce{c-C3H2}
		ratios. Colored regions indicate the observed line ratios drawn
		over the range of densities and temperatures that characterize the observed
		emission, for IRAS~16293-2422 with ALMA (hatched blue) and VLA~1623-2417 (solid orange) with single-dish (SD) and interferometric (I) observations. For \ce{c-C3H2}, the value for IRAS~16293 is from the south position.}
	\label{fig:ratiomodel}
\end{figure*}

\begin{table*} 
	\centering
	\caption{\ce{DCO+} intensities and line ratios, with inferred temperature and density.}
	\label{tab:DCOtemp}
	\begin{tabular}{c c c c c c c c c}
		\hline \hline
		Source & \multicolumn{2}{c}{IRAS~16293-2422} & & \multicolumn{2}{c}{VLA~1623-2417 - ALMA} & & \multicolumn{2}{c}{VLA~1623-2417 - APEX} \\
		\cline{2-3}
		\cline{5-6}
		\cline{8-9}
		Transition & 5--4 (ACA) & 3--2 (SMA) & & 5--4 & 3--2 & & 5--4 & 3--2 \\
		Line width (km s$^{-1}$) & 1.0 & 1.0 & & 0.6 & 0.6 & & 0.7 & 0.7 \\
		Beam (arcsec) & 5.25$\times$2.36 & 4.11$\times$2.45 & & 0.87$\times$0.65 & 0.85$\times$0.54 & & 17.3 & 28.9 \\
		Peak (Jy beam$^{-1}$) & 3.8$\pm$0.6 & 1.9$\pm$0.2 & & 0.29$\pm$0.02 & 0.087$\pm$0.008 & & ... & ... \\
		Peak (K) & 2.9$\pm$0.4 & 4.9$\pm$0.6 & & 4.8$\pm$0.3 & 4.9$\pm$0.5 & & 0.79$\pm$0.1\tablefootmark{a} & 4.8$\pm$0.1 \\ 
		\cline{2-3}
		\cline{5-6}
		\cline{8-9}
		Line ratio & \multicolumn{2}{c}{0.6$\pm$0.1} & & \multicolumn{2}{c}{1.0$\pm$0.1} & & \multicolumn{2}{c}{0.2$\pm$0.03} \\
		Column density (cm$^{-2}$) & \multicolumn{2}{c}{3$\times$10$^{12}$} & & \multicolumn{2}{c}{2$\times$10$^{12}$} & & \multicolumn{2}{c}{2$\times$10$^{12}$} \\
		\ce{H2} density (cm$^{-3}$) & \multicolumn{2}{c}{1--8$\times$10$^{7}$} & & \multicolumn{2}{c}{2--6$\times$10$^{6}$} & & \multicolumn{2}{c}{1--6$\times$10$^{6}$} \\
		Kinetic Temperature (K) & \multicolumn{2}{c}{20--23} & & \multicolumn{2}{c}{30--55} & & \multicolumn{2}{c}{12--19} \\
		$\tau$ & \multicolumn{2}{c}{$<$1} & & \multicolumn{2}{c}{$<$1} & & \multicolumn{2}{c}{$<$1} \\
		\hline
	\end{tabular}
    \tablefoot{\tablefoottext{a}{Peak temperature with beam dilution factor applied to \ce{DCO+} 5--4 (see Appendix~\ref{app:ratios}), taking the beam to be 17.3$\arcsec$, and source 28.9$\arcsec$.}}
\end{table*}

\begin{table*}
	\centering
	\caption{\ce{c-C3H2} inferred parameters and abundance ratio for \ce{c-C3H2}/\ce{C2H}}
	\begin{tabular}{cccccccccc}
		\hline \hline
		Position & \multicolumn{2}{c}{Coordinates} & 9--8/5--4 & 10--9/5--4 & $n_{\ce{H2}}$ & $T_{\rm kin}$ & $N_{\ce{c-C3H2}}$\tablefootmark{a,}\tablefootmark{b} & $N_{\ce{C2H}}$\tablefootmark{a} & \ce{c-C3H2}/\ce{C2H} \\
		& RA & Dec &  &  & cm$^{-3}$ & K & cm$^{-2}$ & cm$^{-2}$ & \\
		\hline
		\multicolumn{10}{c}{IRAS~16293-2422} \\
		\hline
		South & 16:32:22.88 & --24:28:39.78 & 1.4$\pm$0.07 & 1.7$\pm$0.08 & 5--10$\times$10$^{7}$ & 120 -- 155 & 9.3$\times$10$^{13}$ & $\leq$3$\times$10$^{13}$ & $\geq$3.1 \\
		Center & 16:32:22.69 & --24:28:35.16 &  &  & 1--6$\times$10$^{7}$\tablefootmark{c} & 50 -- 120 & 5 -- 7$\times$10$^{12}$ & 2$\times$10$^{14}$ & $\leq$0.035 \\
		North & 16:32:22.55 & --24:28:30.28 &  &  & 1--6$\times$10$^{7}$\tablefootmark{c} & 50 -- 120 & 5 -- 7$\times$10$^{12}$ & 2$\times$10$^{14}$ & $\leq$0.035 \\
		\hline
		\multicolumn{10}{c}{VLA~1623-2417~~ \ce{c-C3H2} ~~6$_{0,6}$--5$_{1,5}$} \\
		\hline
		\ce{c-C3H2}\tablefootmark{e} & 16:26:26.39 & --24:24:30.69 & ... & ... & 5--10$\times$10$^{7}$ & 120--155 & 5.3--6.7$\times$10$^{10}$ & 3$\times$10$^{13}$ & 0.002\tablefootmark{e} \\ 
		\hline
	\end{tabular}
	\label{tab:c3h2temp}
	\\
	\tablefoot{
		\tablefoottext{a}{For IRAS~16293-2422, column densities are for a beam of 0.5$\arcsec$. For VLA~1623-2417, column densities are for a beam of 17.3$\arcsec$, corresponding to the beam of the \ce{C2H} observations.}
		\tablefoottext{b}{An o/p = 3 was used to calculate the total column density of \ce{c-C3H2}.}
		\tablefoottext{c}{Densities are assumed from the model envelope.}
		\tablefoottext{d}{\ce{c-C3H2} $n_{\ce{H2}}$ and $T_{\rm kin}$ parameters taken from the south position of IRAS~16293-2422.}
	    \tablefoottext{e}{Due to the different scales being picked up by the observations, the ratio is not well determined, and is provided here for reference.}}
\end{table*}

\section{Analysis}
\label{sec:analysis}
\ce{DCO+} appears to peak offset from the protostellar positions
bordering the disk-like structures in both sources (Fig.~\ref{fig:i16293mom} and \ref{fig:vla1623mom}). Here we analyze
the peak position first through chemical modelling of the observed emission (Section~\ref{subsec:chemmod}), and then using the line ratios to
constrain the physical structure (temperature, density) and the associated
chemistry (Section~\ref{subsec:dcolineratio}).

To study the physical conditions of the region traced by \ce{c-C3H2}, 
line ratios of the detected transitions are used (Section~\ref{sec:analysis:c3h2}). 
This is combined with \ce{C2H} to obtain the \ce{c-C3H2} / \ce{C2H} abundance ratio and show how the ratio varies with position (Section~\ref{sec:analysis:c2h}).
\ce{c-C3H2} and \ce{C2H} can be produced by the destruction of large hydrocarbons through UV irradiation (top-down chemistry), or through the accumulation of \ce{C} and \ce{H} atoms to form small hydrocarbons (bottom-up). 
Because of the many different formation and destruction pathways, chemical modelling of these two molecules is not included in this work.

\subsection{\ce{DCO+}}
\label{sec:analysis:dco}

\subsubsection{DCO$^+$ distribution}
\label{subsec:chemmod}
In \cite{murillo2015} the distribution of \ce{DCO+} around VLA~1623 was found to be altered 
by the presence of a rotationally supported disk, causing the emission 
to shift inwards along the disk plane but not along other directions.
In this section, the \ce{DCO+} emission around IRAS~16293 is modelled, aiming to
find whether the distribution of \ce{DCO+} in IRAS~16293 is product of the same phenomenon
as observed in VLA~1623.

The \ce{DCO+} chemistry is particularly sensitive to temperature.
To model the observed emission for IRAS~16293 and VLA~1623, a simple steady-state,
analytic chemical network that accounts for the basic reactions 
leading to the production and destruction of \ce{DCO+} is used.  

Since \ce{CO} and \ce{H2D+} are the precursors of \ce{DCO+}, the production of \ce{H2D+} will be the rate-determining reaction in the chemical network, since it will dictate the production of \ce{DCO+}.
The \ce{H2D+} production and destruction reaction is given by
\begin{equation}
\label{eqforward}
\cee{H3+ + HD <=>[][\Delta E] H2D+ + H2}
\end{equation}
where the activation energy $\Delta E \sim$ 220 K in the back reaction
is due to the difference in zero-point energy.
A crucial factor for deuterium chemistry is the ortho-to-para ratio of
\ce{H2} \citep{flower2006,pagani2009}.  This is included in the back
reaction of the chemical network (Eq.~\ref{eqforward}), since it is
here where the distinction has the most significant effect
\citep{murillo2015}. The reactions and parameters for \ce{o-H2} and 
\ce{p-H2} were adapted from \cite{walmsley2004}.  The ortho-to-para 
ratio is set to have a lower limit of 10$^{-3}$ at low temperatures, 
as constrained from observations and models \citep{flower2006}.
The rate coefficient for two-body reaction is expressed as
\begin{equation}
\label{eqratecoeff2b}
k = \alpha \left(\frac{T}{300}\right)^\beta {\rm exp}\left(-\frac{\gamma}{T}\right) ~\rm cm^{3}~s^{-1}
\end{equation}
where $T$ is the temperature of the gas. 
For cosmic ray ionization, important in the generation of \ce{H3+}, 
the rate coefficient is given by
\begin{equation}
\label{eqratecoeff1b}
k = \zeta ~\rm s^{-1}
\end{equation}
where $\zeta$ = 1.26 $\times$ 10$^{-17}$ is the cosmic-ray ionization rate of \ce{H2}.
The reactions and rate coefficients used in this work are the same as those listed in \cite{murillo2015}.

Since \ce{CO} is a
parent molecule of \ce{DCO+}, its abundance will impact the production
of \ce{DCO+}.  The profile of the \ce{CO} abundance is taken to be either constant
or with a drop used to simulate freeze-out.  The drop is set by the
CO sublimation temperature $T_{\rm sub}$ and desorption density $n_{\rm de}$.
These limits dictate the boundary where \ce{CO} is in the gas phase
($T > T_{\rm sub}$) or freezes onto the dust grains ($T < T_{\rm sub}$); and
when the freeze-out time-scales for \ce{CO} are too long ($n <
n_{\rm de}$) compared to the lifetime of the core \citep{jorgensen2005freeze}.
The results of the chemical modelling are passed through RATRAN \citep{hogerheijde2000}, and then synthetic data cubes are generated in order to directly compare with the observations.
Because the \ce{DCO+} emission is weak, the radial profile of the observations is obtained by using a cut with a width that covers the red-shifted peak. The model radial profile, on the other hand, is obtained from a simple cut to the synthetic data cubes since the models are symmetric flat disks.
The network and further details of the model and post-processing are given
in \cite{murillo2015}.

The model requires a density and temperature profile of the source as
a function of radius.  For IRAS~16293, the power-law density and temperature profile
from \cite{crimier2010} is adopted. Two assumptions are made, namely
that the density and temperature profile is centred on IRAS~16293~A  and that
it is the main contributor to the luminosity of the core, consistent 
with the recent analysis of \citet{jacobsen2017}.
\cite{crimier2010} also showed that the emission is dominated by one of the
two components, most likely IRAS~16293 A, rather than being centered between the two sources.
Thus, our assumptions should not introduce major issues in our
modelling.  While there have been several physical profiles derived
for IRAS~16293 \citep[e.g.,][]{schoeier2002}, only one is adopted here since
we alter the density and temperature profiles by an arbitrary
factor, exploring the effects of these parameters on the production of
\ce{DCO+}.  For VLA~1623, we adopt the power-law density and temperature profile from
\cite{jorgensen2002}.  Here again we assume the density and
temperature profile is centred and dominated by VLA~1623 A.  Given that
VLA~1623 B does not contribute much to the line emission nor the continuum,
and that VLA~1623 W is 10$\arcsec$ away, this should not produce
issues in the resulting model.  The temperature and density profile
for VLA~1623 is also altered by an arbitrary factor to study the effect on
\ce{DCO+} production.  The variations in the temperature and density
profiles used in this work are listed in Table~\ref{tab:tests} for
both systems.

For the \ce{DCO+} models, we explore the parameter ranges of $T_{\rm
 sub}$ = 20--40 K, $n_{\rm de}$ = 10$^{5}$--10$^{8}$ cm$^{-3}$ and
$X[\ce{CO}]$ = 10$^{-7}$ -- 10$^{-4}$.  The parameters for the best
by-eye approximation to the observed \ce{DCO+} peak position are
listed in Table~\ref{tab:dcomod} for both systems.
The best approximated model of the \ce{DCO+} 3--2 emission around VLA
\citep{murillo2015} are reproduced here and compared with the results
of \ce{DCO+} toward IRAS~16293.

For both systems we find that the constant \ce{CO} abundance profile
produces a \ce{DCO+} peak further out than where it is observed (gray dashed line in fourth and fifth rows of Fig.~\ref{fig:iras16293chemnet}),
and the peak position does not shift with a change in the abundance \citep{murillo2015}.
The drop \ce{CO} abundance profile produces a peak within the drop
boundaries, $T_{\rm sub}$ and $n_{\rm de}$.  Altering these parameters
changes the shape but not the position of the \ce{DCO+} peak (See \citealt{murillo2015}).  

Since the chemical conditions do not alter the peak position, the physical
structure is examined. The original source density and temperature
profiles for both sources also do not reproduce the position of the
\ce{DCO+} peak. Increasing or
decreasing the density by one order of magnitude, causes the \ce{DCO+} peak to
either shift outwards or remain at a position similar to the unchanged
density profile.  Interestingly, only reducing the temperature
profile by an arbitrary factor together with the drop \ce{CO}
abundance profile, causes the \ce{DCO+} peak to shift inwards for both
systems (orange solid line in Fig.~\ref{fig:iras16293chemnet}).  
A by eye fit of the chemical model to the observations is used 
to constrain the decrease in the temperature profile (Fig.~\ref{fig:iras16293chemnet}). The factor is 
constrained to be 1.5+/-0.2 for VLA 1623 and 3.0+/-0.2 for IRAS 16293. 
This is consistent with
the results found for VLA's \ce{DCO+} 3--2 in \cite{murillo2015},
which explores the physical and chemical parameter space in more
detail.  Thus the observed \ce{DCO+} peak position is produced by a
drop in the temperature along the plane perpendicular to the
outflow(s).  This drop in temperature can be caused by a structure,
such as a disk, which shadows the outer regions, allowing the peak
emission of molecules whose abundance is enhanced in cold gas to move
inwards.

It should be noted, however, that our simple chemical model cannot
fully explain the inner part of the \ce{DCO+} 5--4 emission observed 
toward VLA~1623 A and IRAS~16293 A (orange and black solid lines in the bottom row of Fig.~\ref{fig:iras16293chemnet}).  
\ce{DCO+} 5--4 emission at small radii could be
located in the disk where both cold and warm chemical processes can
contribute to its formation \citep{favre2015,huang2017,salinas2017}.
For warm \ce{DCO+} to form, gas with temperatures up to 70 K \citep{favre2015} are needed. \ce{DCO+} 5--4 is observed further into the disk of VLA~1623 A ($L_{\rm bol}$ = 1 L$_{\odot}$, \citealt{murillo2013a}) than in IRAS~16293 ($L_{\rm bol}$ = 18 L$_{\odot}$, \citealt{jacobsen2017}) due to the lower temperature of the inner disk regions. 
Considering the original source profile which applies to the disk region (top row, solid red line, Fig.~\ref{fig:iras16293chemnet}),
IRAS~16293 A reaches gas temperatures of 70 K at about 1$\arcsec$ or 120 AU, the outer part of the 200~AU disk-like structure \citep{oya2016}. In contrast, VLA~1623 A reaches 70 K at 0.1$\arcsec$ or 12 AU -- i.e., significantly closer to the protostar.

Thus, the distribution of the cold \ce{DCO+} around both VLA~1623 and IRAS~16293 is product 
of the presence of a disk(-like) structure, which causes a drop in
temperature on the envelope gas at the edge of the disk(-like structure), i.e. 
the disk-envelope interface. The presence of the disk(-like) structure
generates an asymmetric temperature profile in the protostellar system.

\subsubsection{Line ratios and implied physical conditions}
\label{subsec:dcolineratio}
Line ratios can provide an independent measure of the temperature of the 
region being traced by a molecule. The ratio of \ce{DCO+} 5--4/3--2 will
provide an independent test of the results obtained with the chemical model
of \ce{DCO+} described in the previous section.

Using RADEX \citep{vandertak2007}, we performed non-LTE excitation and
radiative transfer calculations to constrain the temperature and
density of the regions being traced by comparing the ratios of observed
molecular lines with those calculated by the non-LTE excitation. 
We limit the range of \ce{H2} densities based on the source profile
used for chemical modelling (Table~\ref{tab:tests} and Fig.~\ref{fig:iras16293chemnet})
and the radial position of the emission being modelled.
Using RADEX the column density of the emission was checked to see if it is produced by molecular line emission that is optically thin or thick at \ce{H2} densities of 1--8$\times$10$^{7}$ cm$^{-3}$ for IRAS~16293 and 2--6$\times$10$^{6}$ cm$^{-3}$ for VLA~1623. For the best-fitting column densities of (2-3)$\times$10$^{12}$ cm$^{-2}$ and densities of 10$^{6}$--10$^{8}$ cm$^{-3}$, the \ce{DCO+} 3--2 and 5--4 emission is optically thin in both sources (Table~\ref{tab:DCOtemp}). To produce optically thick lines, column densities of $>$7$\times$10$^{12}$ cm$^{-2}$ would be needed for a temperature of 20 K.
As the emission of both lines are optically thin, the line intensity ratios are not affected by the adopted column density.
All the molecular data files used in this work are
obtained from the Leiden Atomic and Molecular Database 
\citep[LAMDA;][]{schoier2005}. The collisional rate coefficients for \ce{DCO+} are based on the results of \citep{botschwina1993} and \citep{flower1999}.  In order to compare the observed peak
intensities with the results from RADEX, the observed peak intensities are
converted from Jy~beam$^{-1}$ to K using the relation
$T_{\rm mb} = 1.36~{\lambda^{2}}/{\theta^{2}}$$~I_{\rm \nu,obs}$
where $\lambda$ is the wavelength in centimetres of the molecular transition, $\theta$ is the beam of the observations and $I_{\rm \nu,obs}$ is the observed peak intensity in mJy~beam$^{-1}$.

Here, we derive the physical parameters from the \ce{DCO+}
5--4/3--2 ratio for both sources.  Figure~\ref{fig:ratiomodel} shows
the variation of the \ce{DCO+} 5--4/3--2 ratio with \ce{H2} density
and temperature.  For both IRAS~16293 and VLA~1623, the red-shifted peak emission
is considered, since it is the most prominent.  The results for IRAS
and VLA~1623 are compared in Table~\ref{tab:DCOtemp}.  

For IRAS~16293, a ratio \ce{DCO+} 5--4/3--2 = 0.6 $\pm$ 0.1 is obtained 
from the PILS Band 7 observations and
the SMA 230 GHz observations \citep{jorgensen2011}.
Note that for IRAS 16293, we only use the peak intensity from the ACA observations, and thus pick up emission from scales similar to the SMA observations. Thus beam dilution does not need to be taken into consideration. 
We adopt a line width of 1.0 km~s$^{-1}$ and a column density of 3
$\times$ 10$^{12}$ cm$^{-2}$, a value that also reproduces the
observed line intensities. For densities below $10^6$ cm$^{-3}$,
the critical density of the 5--4 transition, the line ratio is
primarily sensitive to density; at higher densities, the ratio becomes
a good temperature probe. According to the density structure presented
in Fig.~ \ref{fig:iras16293chemnet} (top panel), the density at the
peak \ce{DCO+} emission position is higher than the critical density,
so a kinetic temperature between 20 and 23 K can be inferred for IRAS~16293.  
This temperature is consistent with the chemical
modelling of the \ce{DCO+} peak position. 

For VLA~1623, the ALMA 12m array observations provide \ce{DCO+} 5--4/3--2 =
1.0 $\pm$ 0.1. The beam-size of \ce{DCO+} 5--4 (0.87$\arcsec$$\times$0.65$\arcsec$) 
is similar to that of the 3--2 transition
(0.87$\arcsec$$\times$0.54$\arcsec$) and thus no beam dilution factor was
added to the calculation. This line brightness temperature ratio implies a kinetic 
temperature between 30 and 55 K, adopting a column density of 
2 $\times$ 10$^{12}$ and a line width of 0.7 km~s$^{-1}$ to reproduce 
the observed peak intensities. This is higher than expected from the 
chemical modelling of \ce{DCO+}.  The APEX observations are used to 
double check if this is the kinetic temperature of the bulk of the 
\ce{DCO+} emission at the disk-envelope interface.  
The APEX \ce{DCO+} data give a much lower line ratio, 5--4/3--2 = 0.2 
$\pm$ 0.03. 
This line brightness temperature ratio is well reproduced by a kinetic
temperature of 12 to 19 K, in agreement with the chemical model.  It
is likely that the ALMA 12m array observations are picking up both
warm and cold \ce{DCO+} emission in the 5--4 transition, but only cold
\ce{DCO+} in the 3--2 transition.  On the other hand, the APEX
observations are recovering \ce{DCO+} emission from the cold regions
at the edge of the disk and the envelope, but the beam size dilutes
the emission from the inner regions.  This then causes the discrepancy
of derived kinetic temperatures that we obtain from interferometric versus single-dish data. The temperature from
the interferometric data is driven up due to more emission being
detected in the higher transition.

\subsection{\ce{c-C3H2} excitation}
\label{sec:analysis:c3h2}
 
Five transitions of \ce{c-C3H2} are detected towards IRAS~16293.
Temperature is derived from the \ce{c-C3H2} 9--8/5--4 and
10--9/5--4 ratios following the same method as in Section~\ref{sec:analysis:dco}. The collisional rate coefficients for \ce{c-C3H2} are based on \citet{chandra2000}.
The density range is chosen based on the envelope model of IRAS~16293.
Figure~\ref{fig:ratiomodel} shows the the line brightness temperature ratios 
as functions of \ce{H2} density and kinetic
temperature. The ortho-\ce{c-C3H2} molecular file is
used for the RADEX calculations since the 5--4 transition (349.264 GHz) 
presented here is the ortho form (para-\ce{c-C3H2} 5--4 is at 338.204 GHz). 
To convert to the total (ortho + para) \ce{c-C3H2} column density, an o/p ratio of 3 was used.    
Three regions covering the south \ce{c-C3H2} peak and the \ce{C2H}
peaks near the center and north of the map are selected to derive the
temperature and \ce{c-C3H2}/\ce{C2H} column density ratios 
(Fig.~\ref{fig:i16293mapspec}). 
For the position with no detections of either molecule, the peak intensity from within the box used to probe the respective position (Fig.~\ref{fig:i16293mapspec}) is used. These values are listed in Table~\ref{tab:c3h2_c2h_peaks}. 
Table~\ref{tab:c3h2temp} lists the positions.  
The IRAS~16293~A  and B positions are not modelled
due to contamination from other molecular species.

Both \ce{c-C3H2} 9--8/5--4 and 10--9/5--4 ratios are simultaneously fit
for the south position. 
For the center and north positions, a temperature of 50 to 120 K is assumed and the column density is calculated assuming an upper limit of 3 times the rms noise for the peak temperature brightness.
Table~\ref{tab:c3h2temp} lists the derived kinetic temperature from the 
\ce{c-C3H2} ratio together with the assumed \ce{c-C3H2} column densities for 
the regions being traced. 
The total column density for \ce{c-C3H2} is calculated assuming an ortho-to-para ratio of 3.
The peak intensities for each transition are listed in Table~\ref{tab:c3h2_c2h_peaks}
We find that the temperature for the south \ce{c-C3H2} peak, corresponding
to the outflow cavity of IRAS~16293~A , is between 120--155 K.
Comparing the temperatures obtained from \ce{c-C3H2} and \ce{DCO+} line ratios,
it is clear that \ce{c-C3H2} arises from a much warmer region than \ce{DCO+}.
This is linked to the spatial anti-correlation found for these two molecules,
both in our observations and other work (see Sect.~\ref{sec:dis}).

Only one transition of \ce{c-C3H2} is available for VLA~1623, with an
upper level energy (38.6 K) lower than those observed toward IRAS~16293 ($\geq 49$
K).  Thus to obtain an idea of the column densities in the region
traced by \ce{c-C3H2} towards VLA~1623, we adopt the temperature and
density from the \ce{c-C3H2} line ratios towards IRAS
at the south peak.  The \ce{c-C3H2} south peak
of IRAS~16293 is chosen due to the fact that it traces the outflow
cavity, as it does for VLA~1623.  Using the parameters of \ce{c-C3H2} from
the south position of IRAS~16293, the derived column density is 2$\times$10$^{13}$ cm$^{-2}$ for a beam of 1.6$\arcsec$$\times$0.88$\arcsec$,
lower by about a factor of 5 than that found for the south peak
of IRAS~16293 and similar to the column density found for the central
position of the IRAS~16293 map.  
If the beam from the \ce{C2H} observations is considered (17.3$\arcsec$), the derived column density of \ce{c-C3H2} becomes a few 10$^{10}$ cm$^{-2}$ (Table~\ref{tab:c3h2temp}).

\subsection{\ce{c-C3H2} / \ce{C2H} abundance ratio}
\label{sec:analysis:c2h}
The same transitions of \ce{C2H} are observed towards both sources,
with ALMA for IRAS~16293 and with APEX for VLA~1623.  The peak intensities are
listed in Table~\ref{tab:c3h2_c2h_peaks}.  Since \ce{C2H} ratios are
not sensitive to temperature or density given the similar upper eneriges 
$E_{\rm up}$ (Table~\ref{tab:lines}), the method adopted for \ce{DCO+}
and \ce{c-C3H2} cannot be used here.  Instead, the \ce{C2H} column
density is derived by assuming the kinetic temperature and number
density obtained from \ce{c-C3H2} line ratios. Using the same derived temperature and density for \ce{C2H} from \ce{c-C3H2} for the corresponding position is done to probe the parameters of \ce{C2H} if it is tracing the same region as \ce{c-C3H2}.
The results are listed in Table~\ref{tab:c3h2temp}.

The south position in IRAS~16293 presents a column
density of $\leq$3$\times$10$^{13}$ cm$^{-2}$ for \ce{C2H}, about an order of
magnitude lower compared to the centre and north positions which have
a column density of 2$\times$10$^{14}$ cm$^{-2}$.
Table~\ref{tab:c3h2temp} lists the \ce{c-C3H2}/\ce{C2H} column density
ratio for each position.  It must be noted that the
ratio at the south position is a lower limit, whereas for the central
and north position, it is an upper limit.  The differences in
ratios between positions reflect the anti-correlation of the two
molecules in the IRAS~16293 system. Most certainly, the
anti-correlation is not due to critical densities, since the derived
number densities of \ce{c-C3H2} (10$^{7}$ to 10$^{8}$ cm$^{-3}$) at
all points are above the critical densities of both \ce{C2H}
(8$\times$10$^{4}$ to 6$\times$10$^{5}$ cm$^{-3}$) and \ce{c-C3H2}
(2--5$\times$10$^{5}$ cm$^{-3}$).

For VLA~1623, \ce{C2H} column densities are found to be a few times
10$^{13}$ cm$^{-2}$, lower than the peaks of \ce{C2H} detected towards
IRAS~16293.  
The difference in column densities are most likely be due to beam dilution effect from the observations with APEX. Thus the \ce{c-C3H2}/\ce{C2H} ratio is not well determined for VLA1623.
The results are listed in Table~\ref{tab:c3h2temp}.

\subsection{\ce{N2H+} and \ce{N2D+}}
\label{sec:analysis:nitrogen}

For VLA~1623, the ALMA 12m array observations of \ce{N2D+} and \ce{N2H+}
did not detect any emission. Since these molecules are readily
detected in other sources \citep[e.g.,][]{tobin2013}, the cause of
this non-detection is examined. Two cases are explored, extended 
and compact emission concentrated in a 1$\arcsec$ region. 
The details of the analysis are given in Appendix~\ref{app:ratios}.

For the case of extended emission, the predicted \ce{N2H+} 4--3 peak 
intensity is the same as the noise level of our observations, while for the case of the
emission concentrated in 1$\arcsec$ region, the $S/N$ would be about 26.
In a similar manner, the predicted \ce{N2D+} 3--2 peak intensity is
expected to have a marginal detection in our observations for the extended emission
case, and a $S/N$ = 40 for the compact emission case. Thus, we should have detected
both molecules in our observations if they arose from a compact structure.

\begin{figure}
	\centering
	\includegraphics[width=0.95\columnwidth]{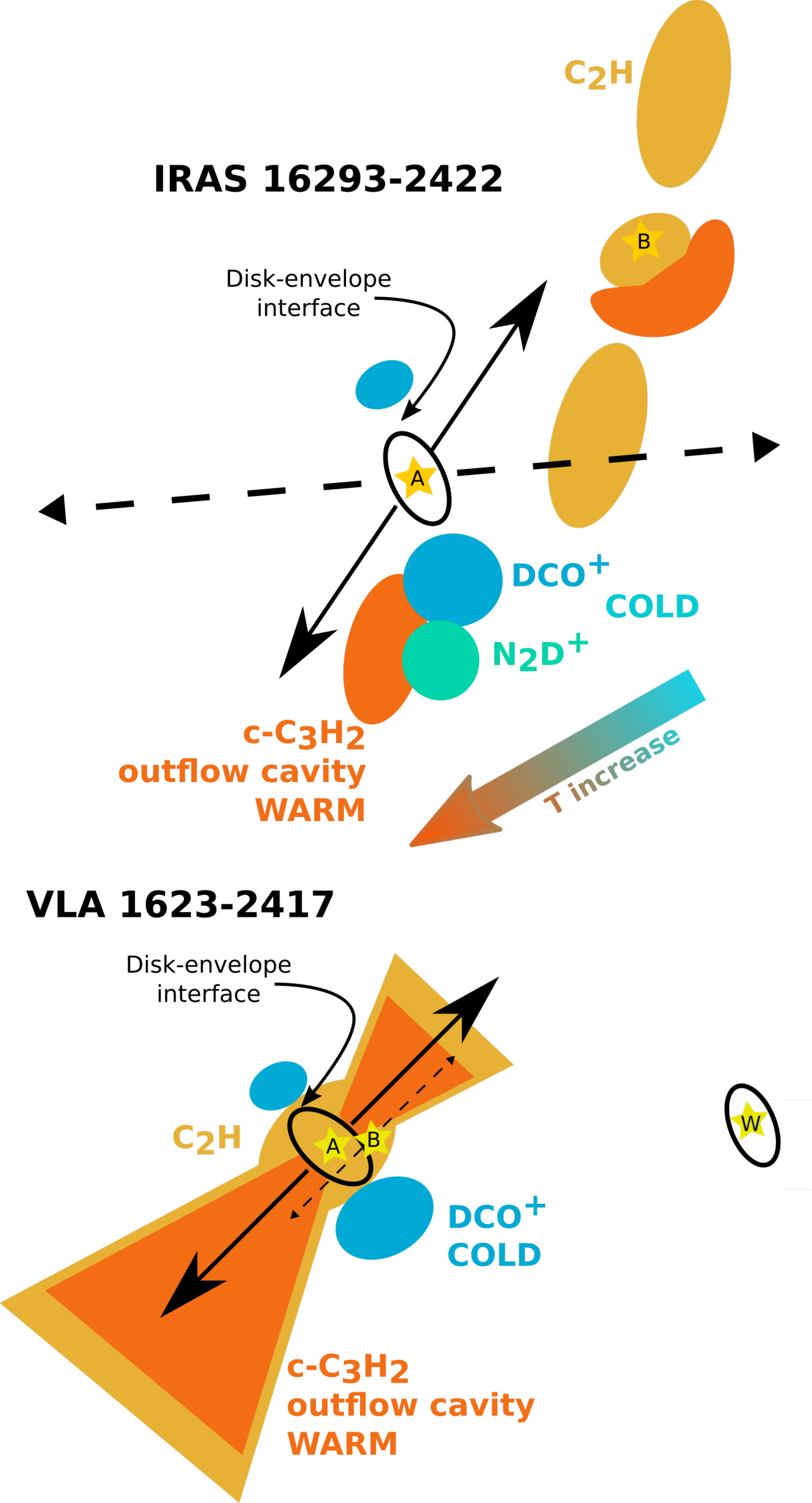}
	\caption{Cartoon showing the distribution of the molecules studied in this work towards both sources. The solid and dashed lines show outflow directions. Black ellipses indicate disk structures. \ce{C2H} toward VLA~1623A is observed with APEX, thus the distribution is expected based on the physical conditions of VLA~1623, instead of direct mapping.}
	\label{fig:cartoon}
\end{figure}

\section{Discussion}
\label{sec:dis}

\subsection{Comparison of IRAS~16293-2422 and VLA~1623-2417}
The chemical structure of both systems is compared in this section.
A cartoon of their structure is shown in Fig.~\ref{fig:cartoon}.

The \ce{DCO+} peak position in both sources is well described by a drop in the temperature profile. 
This drop can be explained by the presence of a disk which shadows the envelope, causing the \ce{DCO+} emission to move inward, closer to the source along the disk plane.
For VLA~1623, the ALMA \ce{DCO+} 5--4 observations are picking up emission coming from both the cold envelope at the edge of the disk (disk-envelope interface) and the warmer parts of the disk.

\ce{c-C3H2} traces the outflow cavity wall of IRAS~16293~A  and VLA~1623 A.
Due to \ce{c-C3H2} having both top-down and bottom up chemistry, it was not studied using a chemical model.
For VLA~1623 A, \ce{c-C3H2} traces the full outflow cavity (extending out to 3$\arcsec$), whereas for IRAS~16293~A  only one side of the south outflow cavity wall is observed.
It is possible that the \ce{c-C3H2} emission is product of UV radiation from the central source, and the warmer temperatures found in the outflow cavity.
UV radiation liberates atomic carbon which leads to gas-phase formation of small hydrocarbons.
Chemical models of the outflow cavity walls show that \ce{c-C3H2} is initially concentrated around the outflow wall, and progressively moves to the disk plane as the cavity widens with age, irradiating more envelope material \citep{maria2015}.
Higher temperatures could accelerate certain chemical processes in the protostellar envelope, while movement of material through outflows, rotation and infall could bring dust with \ce{c-C3H2} precursors (e.g., \ce{CH4}, \ce{C2H2}) closer to areas where they can be sublimated and thus enhance the \ce{c-C3H2} gas.
Hence, the differences in the spatial distributions of \ce{c-C3H2} in IRAS~16293~A ($L_{\rm bol}$ $>$ 18 L$_{\odot}$, \citealt{jacobsen2017}) and VLA~1623 A ($L_{\rm bol}$ $\sim$1$L_{\odot}$) could be the product of age, luminosity, or core dynamics. 
In any case, the presence of \ce{c-C3H2} emission is due to the temperature of the region where it is observed.

\ce{C2H} does not present similar distributions in IRAS~16293 and VLA~1623.
For VLA~1623, the single-dish observations of \ce{C2H} showed no emission was detected in the off positions, thus we expect that \ce{C2H} peaks around VLA~1623 A. Given the observations, low temperatures in the envelope of VLA~1623 A as well as the envelopes of VLA~1623 B and W, and the formation mechanism of the two molecules, it might be possible that \ce{C2H} is spatially correlated with \ce{c-C3H2} on $\sim$15$\arcsec$ scales. However, interferometric observations are needed to verify if both molecules are correlated toward VLA~1623. 
On larger scales, part of the \ce{C2H} could also come from the envelope, given that it shows similar broadening with \ce{DCO+} \citep{lindberg2017}.

For IRAS~16293, the observed \ce{C2H} is found in the region one beam away from IRAS~16293~B, but does not show relation with the position or outflow of IRAS~16293~A, nor with the dust ridge connecting both sources.
Beyond IRAS~16293~B, \ce{C2H} and \ce{c-C3H2} are not spatially correlated (Fig.~\ref{fig:i16293mapspec} and \ref{fig:i16293spec}).
This anti-correlation is unexpected from chemical models (e.g. \citealt{gerin2011,maria2015,guzman2015}) or observations (see Sect.~\ref{sec:diss:comp}).
The \ce{c-C3H2}/\ce{C2H} ratio is expected to be lowered with age, that is as the protostar evolves (O. Sipil{\"a}, private communication).
If the different distributions of \ce{c-C3H2} and \ce{C2H} are product of age, it would suggest that IRAS~16293 and VLA~1623 are young; however, lowering the ratio with age does not explain the anti-correlation observed within IRAS~16293.

A possible explanation may be top-down chemistry producing small hydrocarbons through UV destruction of large aromatic molecules as inferred for lower density PDRs \citep{guzman2015}, producing different amounts of the two molecules.
Another possibility for the anti-correlation might be explained by the destruction of \ce{C2H} in reactions with sulfur, nitrogen, oxygen or carbon chains \citep{sakai2013}.
The full Band 7 spectrum is examined at the \ce{c-C3H2} peak position (Fig.~\ref{fig:fullspec1}--~\ref{fig:fullspec3}) in order to examine whether products of \ce{C2H} reactions are present. 
\ce{C2S}, product of \ce{C2H} reacting with sulfur \citep{drozdovskaya2018}, is not detected.
In fact, little else is observed in the dense gas south of IRAS~16293~A.
Apart from \ce{c-C3H2}, only \ce{H2CS} \citep{drozdovskaya2018} and a few common species like \ce{HCO+}, \ce{H2CO} and \ce{CH3OH} are detected, which is unexpected given that outflow cavity would be irradiated and encourage chemical complexity \citep{maria2015}.
Reactions of \ce{C2H} with carbon chains \ce{C_{n}} would cause production of \ce{C_{n+2}} and hydrogen, rendering these products undetectable due to lack of dipole moments. 
Overall, the strong anti-correlation of these molecules remains a chemical puzzle.

\ce{N2D+} is detected 7$\arcsec$ south of IRAS~16293~A  with the SMA (Fig.~\ref{fig:i16293SMA}), bordering the \ce{DCO+} emission \citep{jorgensen2011}.  
In contrast, \ce{N2D+} and \ce{N2H+} are not detected with ALMA observations towards VLA~1623.
Single-dish observations show an offset of about 60$\arcsec$ between the position of VLA~1623 and the peak of \ce{N2H+} and \ce{N2D+} (\citealt{difrancesco2004}; \citealt{andre2007}; \citealt{liseau2015,punanova2016}; \citealt{favre2017}). 
Line ratios of \ce{N2H+} and \ce{N2D+} from single-dish observations were investigated in Section~\ref{sec:analysis:nitrogen} and Appendix~\ref{app:ratios}, where it was found that ALMA would detect only compact emission. Given the lack of ALMA detection of both molecules, \ce{N2D+} and \ce{N2H+} trace extended emission outside the envelope of VLA~1623 \citep{liseau2015}. 

The presence of \ce{N2D+} in IRAS~16293, but not in VLA~1623, could be product of temperature differences.
\ce{N2} can be frozen out onto dust grains at temperatures below 20 K \citep{bisschop2006}, a scenario also pointed out by \cite{difrancesco2004}. 
While other nitrogen-bearing molecules such as \ce{CN}, \ce{HCN}, \ce{HNC} and \ce{NO} can form in the gas and on grain surfaces, \ce{N2H+} and \ce{N2D+} only form in the gas phase if \ce{N2} gas is present. 
This scenario is further supported by the low temperatures found for \ce{DCO+}. 
For VLA~1623 A, \ce{DCO+} has $T_{\rm kin}$ = 17 K, and the chemical modelling suggests dust temperatures between 11 to 16 K for where \ce{DCO+} peaks. 
This would indicate that further out, the temperature is even lower.
In addition, at densities below $\sim$10$^{4}$ cm$^{-3}$ (two orders of magnitude below what is derived from \ce{DCO+} for VLA~1623), the dust and gas temperatures decouple, and without any additional external pressure, the gas temperature drops down to 10 K \citep{galli2002,evans2001}, which could cause \ce{N2H+} and \ce{N2D+} to recombine onto the dust grains or the precursor \ce{N2} to freeze-out.
In contrast, \ce{DCO+} south of IRAS~16293~A  indicates $T_{\rm kin}$ = 20 to 23 K for the gas, and dust temperatures between 17 to 19 K from chemical modelling, evidencing that the envelope of IRAS~16293~A  is warmer than that of VLA~1623 \citep{jacobsen2017}.

\ce{N2H+} and \ce{N2D+} are thought to be tracers of evolutionary stage \citep{emprechtinger2009}, as well as of the \ce{CO} snowline \citep{jorgensen2004,anderl2016,merel2017}.  
These assumptions break down for very cold envelopes of embedded protostars, like that of VLA~1623.  
Given that some starless cores do show \ce{N2H+} and \ce{N2D+} (e.g. \citealt{crapsi2005,tobin2013}), including the starless cores north of VLA~1623 \citep{difrancesco2004,friesen2014}, it cannot be said that the cold envelope itself is an indicator of evolutionary stage.  
It may be possible that the ridge of material north of VLA~1623, which contains the starless cores, is being heated somehow from the side, but VLA~1623 is being shielded and thus much colder \citep{difrancesco2004,bergman2011,friesen2014}.  
VLA~1623 A itself is certainly heating up the disk and outflow cavity, evidenced by \ce{DCO+} 5--4 emission on the disk and the presence of \ce{c-C3H2} and \ce{C2H}, but on much smaller scales ($<$100~AU) than in IRAS~16293 because of its lower luminosity.

\subsection{Comparison with starless cores and low-mass protostars}
\label{sec:diss:comp}
In this section, IRAS~16293 and VLA~1623 are placed in the big picture of star formation.
For this reason, the two systems described in the previous section are compared with observations of starless cores, embedded low-mass protostars and disks found in the literature.
In addition, the multiplicity of the systems is also taken into consideration.

The starless core L1544 exhibits \ce{c-C3H2} close to the dense cloud core center and away from cold regions traced by \ce{DCO+} \citep{spezzano2016a,spezzano2016b}.
This points to an anti-correlation between the chemistry traced by \ce{DCO+} and that by \ce{c-C3H2}, which is present in both IRAS~16293 and VLA~1623.
In the system NGC1333 IRAS4, \ce{C2H} is observed to peak on-source toward each component, including the starless core IRAS4C, which has the strongest emission \citep{koumpia2016,koumpia2017}.
In the young embedded object IRAS15398, \ce{C2H} traces the red- and blue-shifted outflow cavity \citep{jorgensen2013}.
In contrast, L1527 presents both \ce{C2H} and \ce{c-C3H2} in the envelope and disk component, with enhancements at the centrifugal barrier \citep{sakain2010,sakai2014,sakai2016}, but no emission along the outflow cavity.
The spatial distribution of \ce{C2H} and \ce{c-C3H2} is similar in L1527, with the emission from \ce{c-C3H2} being more compact than that of \ce{C2H}.
In Oph-IRS67, \ce{C2H} and \ce{c-C3H2} exist in the same region, although the spatial extent is not the same \citep{artur2018}.
In the protoplanetary disk TW Hya, \ce{C2H} and \ce{c-C3H2} are found to reside in the disk, bordering the millimeter dust, with both molecules showing an identical spatial distribution \citep{bergin2016}.
The \ce{c-C3H2} and \ce{C2H} distribution toward VLA~1623 is consistent with that observed in other protostellar systems; however, for IRAS~16293 the lack of correlation between the two molecules is still a puzzle, since no other low-mass protostar or starless core reported in the literature at present presents this situation.

In NGC1333 SVS13, \ce{N2H+} is detected around 2 of the 4 components of the system \citep{chen2009}.
From the system, SVS13B and SVS13C are Class 0 protostars, but the first has \ce{N2H+} emission while the second does not.
Thus, the uneven distribution of material is not related to the evolutionary stage, but instead is most likely related to the varying envelope temperature.

Several of the systems mentioned above are multiple protostars, as are IRAS~16293 and VLA~1623. 
The chemical structure is found to not be homogeneous among the individual components of these systems. 
This would suggest that the components of wide multiple protostellar systems have no effect on the chemistry of each other. For close multiple protostellar systems, the only case shown here is that of VLA1623 A and B, which have a separation of $\sim$200~AU (based on the disk radius of VLA1623 A and a lack of disturbance of the disk by VLA1623 B) and show different chemical structures. However, these are only two cases, and more observations of multiple protostellar systems are needed to further understand the effect of companions on chemical structure.

\subsection{Comparison with diffuse clouds, PDRs and intermediate to high-mass protostars}
Looking to compare what structures are common throughout the interstellar medium, we compare the distributions found in IRAS~16293 and VLA~1623 with diffuse clouds and PDRs. 
Furthermore, given that IRAS~16293 is much warmer than VLA~1623, it is also compared to high-mass protostars.

Towards the Horsehead nebula PDR, \ce{DCO+} is observed far from the irradiated edge of the region, with no emission at the edge \citep{guzman2015}.  
The spatial anti-correlation between \ce{DCO+} and \ce{c-C3H2} or \ce{C2H} suggests a temperature effect, as found for IRAS~16293 and VLA~1623, highlighting that \ce{DCO+} is a really good tracer of cold regions.

\ce{c-C3H2} and \ce{C2H} show close correlation in spatial distribution towards a number of PDRs, including the Orion Bar \citep{pety2007,wiel2009,nagy2015} and the Horsehead Nebula \citep{cuadrado2015,guzman2015}, with both molecular species sitting at the irradiated, and thus warmer, edge of the region.  
In addition, a tight correlation between \ce{c-C3H2} and \ce{C2H} in diffuse clouds has been found \citep{lucas2000,gerin2011,liszt2012}. 
The column density \ce{c-C3H2}/\ce{C2H} ratios calculated toward IRAS~16293 in the center and north positions (\ce{c-C3H2}/\ce{C2H} $\leq$ 0.035) reflect the values found for diffuse clouds (\ce{c-C3H2}/\ce{C2H} = 0.048; \citealt{lucas2000,liszt2012}) and the envelope of L1527 (\ce{c-C3H2}/\ce{C2H} = 0.035 -- 0.06; \citealt{sakai2014}).

In high-mass star-forming regions, \ce{c-C3H2} and \ce{C2H} tend to be strongly correlated, with both lines presenting similar spatial distributions \citep{pilleri2013,mookerjea2012,mookerjea2014}.  
In contrast, toward IRAS 20343+4129, \ce{c-C3H2} and \ce{C2H} show an anti-correlation around the outflow cavity walls of IRS 1 \citep{fontani2012} but not around the UC HII-region of IRS 3. The anti-correlation is explained to possibly be product of the gas density, with an enhancement of \ce{C2H} located in the regions with denser gas.
Thus, the warmer envelope of IRAS~16293 does not provide a solution to the puzzle of why \ce{c-C3H2} and \ce{C2H} are anti-correlated in this system, but the difference in density, possibly caused by the outflow direction shift, might provide a possible explanation for the anti-correlation of the two molecules.
Chemical processes could also be playng a role in the anti-correlation of \ce{c-C3H2} and \ce{C2H} toward IRAS~16293.

\section{Conclusions}
\label{sec:conc}
In this work, we present ALMA, SMA and APEX observations of \ce{DCO+},
\ce{c-C3H2}, \ce{C2H}, \ce{N2H+} and \ce{N2D+} towards IRAS~16293-2422
and VLA~1623-2417, both multiple protostellar systems in $\rho$
Ophiuchus.  The spatial distribution of each molecule is compared for
both systems.  \ce{DCO+} is studied using a simple analytic chemical
network coupled with radiative transfer modelling, detailed in
\cite{murillo2015}, in order to determine the conditions leading to
the observed peak position.  Non-LTE molecular excitation and
radiative transfer modelling of the observed line brightness temperature ratios is done
to derive physical parameters of the regions being traced by the
molecules.  Finally, the observations and results of VLA~1623-2417 and
IRAS~16293-2422 are compared, both between the two sources and other
objects, ranging from low- to high-mass protostars, diffuse clouds and
PDRs, in order to understand what structures are common.

From this work, we extract the following key points:
\begin{enumerate}
\item Temperature is a controlling factor of the chemical structure of a protostellar system. Disks can alter the temperature of the envelope, while UV heating can encourage the start of chemical processes in the outflow cavity.
\item An asymmetric \ce{DCO+} structure is a good tell-tale sign for the presence of a disk, since a disk shadows the envelope at its edge, lowering the temperature and causing \ce{DCO+} to move inwards only along the disk plane.
\item \ce{c-C3H2} traces the outflow cavity of IRAS~16293-2422 and VLA~1623-2417, but shows no disk component for either source.
\item Despite both VLA~1623-2417 and IRAS~16293-2422 being low-mass Class 0 embedded objects, their structure and chemical richness varies considerably, with VLA~1623-2417 being line poor. Its much lower luminosity, and consequently lower temperatures, coupled with a large cold disk, are likely at the root of this difference. 
\end{enumerate}

Although only two sources are studied in this work and some results in the literature, there is evidence pointing to a lack of correlation between multiplicity and the chemical structure of the envelope of these systems, at least in the embedded phase.
Nevertheless, multiple systems do provide an interesting way to compare the structure with similar conditions. It would be possible, however, that the heating from companion sources would affect the chemistry as the envelope clears.
Further comparison of embedded multiple protostellar systems is needed to confirm these results.

\begin{acknowledgements}
	The authors thank O. Sipil{\"a}, P. Caselli, S. Spezzano, N. Sakai and C. Favre for helpful discussions on chemistry.
	We are grateful to the APEX staff for support with the observations of VLA1623.
	This paper made use of the following ALMA data: ADS/JAO.ALMA 2011.0.00902.S, 2013.1.01004.S and 2013.1.00278.S. ALMA is a partnership of ESO (representing its member states), NSF (USA), and NINS (Japan), together with NRC (Canada) and NSC and ASIAA (Taiwan), in cooperation with the Republic of Chile. The Joint ALMA Observatory is operated by ESO, AUI/NRAO, and NAOJ. The 2011.0.00902.S data was obtained by N.M.M. while she was a Master student at National Tsing Hua University, Taiwan, under the supervision of S.P.L. 
	Astrochemistry in Leiden is supported by the European Union A-ERC grant 291141 CHEMPLAN, by the Netherlands Research School for Astronomy (NOVA), by a Royal Netherlands Academy of Arts and Sciences (KNAW) professor prize.
	The group of JKJ acknowledges support from the European Research Council (ERC) under the European Union's Horizon 2020 research and innovation programme (grant agreement No~646908) through ERC Consolidator Grant ``S4F''. Research at Centre for Star and Planet Formation is funded by the Danish National Research Foundation.
	MND acknowledges the financial support of the Center for Space and Habitability (CSH) Fellowship and the IAU Gruber Foundation Fellowship. 
\end{acknowledgements}

\bibliographystyle{aa}
\bibliography{simple_organics.bib}

\begin{appendix}

\section{Interferometric Observations}
\label{app:uv}
Table~\ref{tab:uv} lists details of the interferometric observations used in this work for both systems, IRAS 16293 and VLA 1623.
For VLA 1623, the Cycle 2 Band 6 12m array observations have both long (C35-5) and short (C34-1) baselines. The short baselines were observed to bridge the baseline ranges between the ACA and 12m observations.
The \ce{c-C3H2} observations toward VLA 1623 presented in this work was detected with the 12m array short baselines but not the long baselines, suggesting that the emission arises from regions larger than $\sim$0.5$\arcsec$.

\begin{table*}
	\centering
	\caption{Interferometric observations}
	\begin{tabular}{ccccccc}
	\hline \hline
	Dataset & Frequency & UV-baseline range & Beam & Largest scale & Field of View & Noise \\
	 & GHz & k$\lambda$ & $\arcsec$ & $\arcsec$ & $\arcsec$ & mJy beam$^{-1}$ \\
	\hline
	\multicolumn{7}{c}{IRAS~16293-2422} \\
	\hline
	Combined 12m+ACA & 329.1 -- 362.8 & 10 $\sim$ 352 & 0.5$\times$0.5 & 13 & 19.5 & 7--10 \\
	ACA & 329.1 -- 362.8 & 10 $\sim$ 60 & 5.25$\times$2.36 & 13 & 60 & 440 \\ 
	SMA Compact & 215.6--227.6 & 6 $\sim$ 91 & 5.5$\times$3.2 & 20 & 64 & 240 \\
	\hline
	\multicolumn{7}{c}{VLA 1623-2217} \\
	\hline
	Cycle 0 12m & 216.4 -- 232.2 & 25 $\sim$ 310 & 0.85$\times$0.56 & 2.5 & 28 & 4--8 \\
	Cycle 2 Band 6 12m C35-5 & 215.5 -- 221.4 & 18 $\sim$ 791 & 0.45$\times$0.25 & 1.9 & 28 & 7--9 \\
	Cycle 2 Band 6 12m C34-1 & 215.5 -- 221.4 & 11 $\sim$ 253 & 1.6$\times$0.88 & 5.3 & 28 & 20 \\
	Cycle 2 Band 7 12m & 359.0 -- 372.7 & 18 $\sim$ 420 & 0.87$\times$0.54 & 3.1 & 17.2 & 26-95 \\ 
	\hline
	\end{tabular}
\label{tab:uv}
\end{table*}

\section{Additional \ce{c-C3H2} and \ce{C2H} spectra}
\label{app:c2h}
\begin{figure*}
	\centering
	\includegraphics[width=0.96\textwidth]{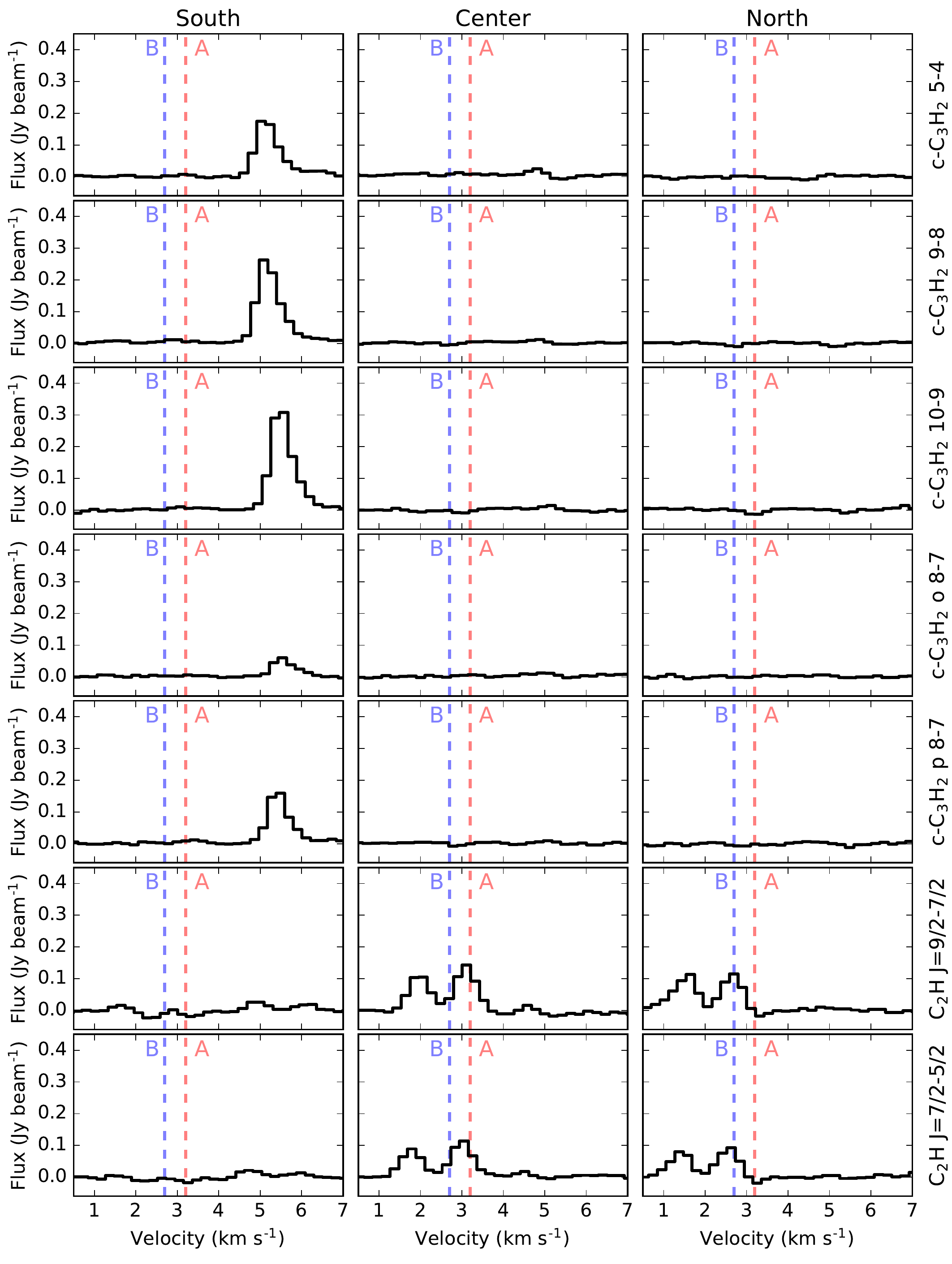} %
	\caption{IRAS~16293-2422 ALMA spectra for all transitions of \ce{c-C3H2} and \ce{C2H} for each of the south, center and north positions shown in Fig.~\ref{fig:i16293mapspec}. The systemic velocities of sources A and B are marked with the vertical dashed lines. The anti-correlation of \ce{c-C3H2} and \ce{C2H} is also evident in the spectra shown here.}
	\label{fig:i16293spec}
\end{figure*}

The spectra for all transitions of \ce{c-C3H2} and \ce{C2H} toward the south, center and north positions of IRAS~16293 are shown in Fig.~\ref{fig:i16293spec}. The systemic velocity of IRAS~16293 A and B are marked on the spectra with dashed lines. The anti-correlation of the two molecules is evident from the spectra, as well as a slight velocity shift in \ce{C2H} between the center and north positions.  

\section{Peak intensities and line ratios}
\label{app:ratios}
The observed peak intensities of \ce{C2H} and \ce{c-C3H2} are listed in this appendix (Table~\ref{tab:c3h2_c2h_peaks}).
The peak intensities are used in Sect.~\ref{sec:analysis:c3h2} and \ref{sec:analysis:c2h}. 
In addition, the detailed calculation of expected peak intensities for \ce{N2H+} and \ce{N2D+} are also summarized here (Table~\ref{tab:nitrogen}).

Single-dish observations of low-$J$ transitions of \ce{CN}, \ce{HCN} and \ce{HNC} show strong detections relative to other embedded systems \citep{jorgensen2004}.
Our recent APEX observations also detected \ce{NO} towards VLA~1623.
However, single-dish observations of \ce{NH3} \citep{wootten1994,liseau2003} and \ce{N2H+} \citep{liseau2015,punanova2016} indicate that these molecules have very low abundances at the position of VLA~1623.  
\ce{N2D+} also exhibits the same behaviour \citep{punanova2016}.  
Furthermore, {\it Herschel} observations of high-$J$ \ce{N2H+} towards VLA~1623 (\citealt{liseau2015}; \citealt{favre2017}) show that the molecule is detected up to the $J$ = 6--5 transition peaking at $\sim$0.1 K km~s$^{-1}$ but the emission is extended.  
The observed parameters of \ce{N2H+} and \ce{N2D+} from \cite{punanova2016} and \cite{favre2017} are listed in Table~\ref{tab:nitrogen}. 

Using the observed transitions of \ce{N2H+} and \ce{N2D+}, we derive density and excitation temperature with the method described in Sect.~\ref{sec:analysis:dco}. 
The different beam sizes of the observations require a beam dilution correction factor that is given by $T_{\rm mb}'~=~T_{\rm mb,obs}~\frac{\Omega_{\rm beam}}{\Omega_{\rm source}}$, where $T_{\rm mb}'$ and $T_{\rm mb,obs}$ are the corrected and observed main beam temperature, respectively, $\Omega_{\rm beam}$ is the solid angle of the single-dish beam and $\Omega_{\rm source}$ is the solid angle subtended by the source.  
We assume the emission is concentrated in the region of the smaller beam, which would be of 26.5$\arcsec$ for
\ce{N2H+} and 16.3$\arcsec$ for \ce{N2D+}.

To compare with our ALMA observations, the expected peak for \ce{N2H+} 4--3 and \ce{N2D+} 3--2 is derived from the observations of \cite{punanova2016} and \cite{favre2017}.
The \ce{N2H+} molecular data file without hyperfine structure is used to calculate the kinetic temperature of both molecules.
The collisional rate coefficients for \ce{N2H+} are taken to be the same as \ce{HCO+} \citep{botschwina1993} and extrapolated \citep{schoier2005}.
Since LAMDA does not have a molecular data file for \ce{N2D+}, the data file for \ce{N2H+} is used, selecting the corresponding transition rather than frequency.
For the predicted peak emissions for \ce{N2H+} 4--3 and \ce{N2D+} 3--2, two cases are examined: i) the observed emission is evenly distributed in the single-dish beam (i.e., beam filling factor = 1) and ii) the emission is concentrated in a 1$\arcsec$ region (i.e. beam filling factor $<$ 1). 
The second case introduces a beam dilution correction.  
The results of the calculation are listed in Table~\ref{tab:nitrogen}.

The kinetic temperature and number density obtained in our calculations ($\sim$11 K, $\sim$10$^{7-9}$ cm$^{-3}$) are slightly higher than those previously reported (7.7 K, 10$^{6}$ cm$^{-3}$; \citealt{punanova2016}).
The column densities, however, are similar to those reported in \cite{punanova2016}. For \ce{N2H+}, our results are also consistent with those reported in \cite{liseau2015}. 

Using the physical parameters obtained from the \ce{DCO+} 5--4/3--2 ratio (Sect.~\ref{sec:analysis:dco} and Table~\ref{tab:DCOtemp}), we calculate a column density of 1.5--2$\times$10$^{13}$ cm$^{-2}$ for \ce{N2D+}.  
If instead we use the physical parameters obtained from the \ce{N2D+} observations towards VLA~1623, we find a column density of 4--5$\times$10$^{13}$ cm$^{-2}$ for \ce{N2D+} 3--2 towards IRAS~16293.  
For both sets of parameters, the column density is higher by one order of magnitude in comparison to the \ce{N2D+} toward VLA~1623.

\begin{table*}
\centering
\caption{\ce{c-C3H2} and \ce{C2H} peak intensities.}
\begin{tabular}{ccccccccccc}
\hline \hline
Molecules & \multicolumn{5}{c}{\ce{c-C3H2}} & & \multicolumn{4}{c}{\ce{C2H}} \\
\cline{2-6}
\cline{8-11}
Transition & 5--4 & 9--8 & 10--9 & 8$_{2,6}$--7$_{3,5}$ & 8$_{3,6}$--7$_{2.5}$ & & J=9/2--7/2 F=5--4 & J=9/2--7/2 F=4--3 & J=7/2--5/2 F=4--3 & J=7/2--5/2 F=3--2 \\
\hline
\multicolumn{11}{c}{IRAS~16293-2422 (rms = 10 mJy~beam$^{-1}$)} \\
\hline
South\tablefootmark{a} & 240 & 410 & 350 & 90 & 200 & & 30 & 30 & 30 & 30 \\
Centre\tablefootmark{a} & 30 & 30 & 30 & 30 & 30 & & 170 & 140 & 140 & 110 \\
North\tablefootmark{a} & 30 & 30 & 30 & 30 & 30 & & 150 & 130 & 110 & 110 \\
\hline
\multicolumn{11}{c}{VLA~1623-2417 - APEX (rms = 80 mK)} \\
\hline
A & ... & ... & ... & ... & ... & & 960 & 616 & 680 & 342 \\
\hline
\end{tabular}
\tablefoot{
\tablefoottext{a}{For the positions with non-detections, 3 times the rms noise is used for the calculations.}
}
\label{tab:c3h2_c2h_peaks}
\end{table*}

\begin{table*}
\centering
\caption{\ce{N2H+} and \ce{N2D+} parameters for VLA~1623-2417.}
\begin{tabular}{cccccc}
\hline \hline
& \multicolumn{5}{c}{Observed} \\
\cline{2-6}
Parameter & \ce{N2H+} 1--0\tablefootmark{a} & \ce{N2H+} 6--5\tablefootmark{b} && \ce{N2D+} 1--0\tablefootmark{a} & \ce{N2D+} 2--1\tablefootmark{a} \\
\hline
$T_{\rm mb}$ (K) & 3.3$~\pm~$0.1 & 0.2$~\pm~$0.1 && 0.57$~\pm~$0.1 & 1.5$~\pm~$0.2 \\
$\Delta v_{\rm single~dish}$ (km~s$^{-1}$) & 0.56$~\pm~$0.005 & 0.7$~\pm~$0.2 && 0.55$~\pm~$0.032 & 0.55$~\pm~$0.009 \\
$\eta_{\rm mb}$ & 0.95 & 0.63 && 0.95 & 0.94 \\
$\Omega_{\rm beam}$ ($\arcsec$) & 26.5 & 36 && 32.1 & 16.3 \\
\cline{2-3}
\cline{5-6}
Ratio & \multicolumn{2}{c}{0.06 $\pm$ 0.03} && \multicolumn{2}{c}{2.6 $\pm$ 0.6} \\
Column density (cm$^{-2}$) & \multicolumn{2}{c}{1.3$\times$10$^{13}$} && \multicolumn{2}{c}{1.8$\times$10$^{12}$} \\
\ce{H2} density (cm$^{-3}$) & \multicolumn{2}{c}{5$\times$10$^{7}$ -- 7$\times$10$^{9}$} && \multicolumn{2}{c}{5$\times$10$^{7}$ -- 7$\times$10$^{9}$} \\
$T_{\rm ex}$ (K) & \multicolumn{2}{c}{11 -- 12} && \multicolumn{2}{c}{11 -- 12} \\
\hline
$\Omega_{\rm source}$ = $\Omega_{\rm beam}$ & \multicolumn{2}{c}{\ce{N2H+} 4--3} && \multicolumn{2}{c}{\ce{N2D+} 3--2} \\
\hline
$I_{\rm \nu,predicted}$ (mJy~beam$^{-1}$) & \multicolumn{2}{c}{99} && \multicolumn{2}{c}{22} \\
$\Omega_{\rm beam,obs}$ ($\arcsec$) & \multicolumn{2}{c}{0.85} && \multicolumn{2}{c}{0.76} \\
$\sigma_{\rm obs}$ (mJy~beam$^{-1}$) & \multicolumn{2}{c}{94.9} && \multicolumn{2}{c}{8.58} \\
$S/N$ & \multicolumn{2}{c}{1} && \multicolumn{2}{c}{2.6} \\
\hline
$\Omega_{\rm source}~=~1\arcsec$ & \multicolumn{2}{c}{\ce{N2H+} 4--3} && \multicolumn{2}{c}{\ce{N2D+} 3--2} \\
\hline
$I_{\rm \nu,predicted}$ (mJy~beam$^{-1}$) & \multicolumn{2}{c}{2631} && \multicolumn{2}{c}{364} \\
$\Omega_{\rm beam,obs}$ ($\arcsec$) & \multicolumn{2}{c}{0.85} && \multicolumn{2}{c}{0.76} \\
$\sigma_{\rm obs}$ (mJy~beam$^{-1}$) & \multicolumn{2}{c}{94.9} && \multicolumn{2}{c}{8.58} \\
$S/N$ & \multicolumn{2}{c}{27} && \multicolumn{2}{c}{42} \\
\hline
\end{tabular}
\tablefoot{
\tablefoottext{a}{IRAM 30m observations from \cite{punanova2016}.}
\tablefoottext{b}{\textit{Herschel} observations from \cite{favre2017}}
}
\label{tab:nitrogen}
\end{table*}

\section{PILS full spectra}
\begin{figure*}
\centering
\includegraphics[width=0.98\textwidth]{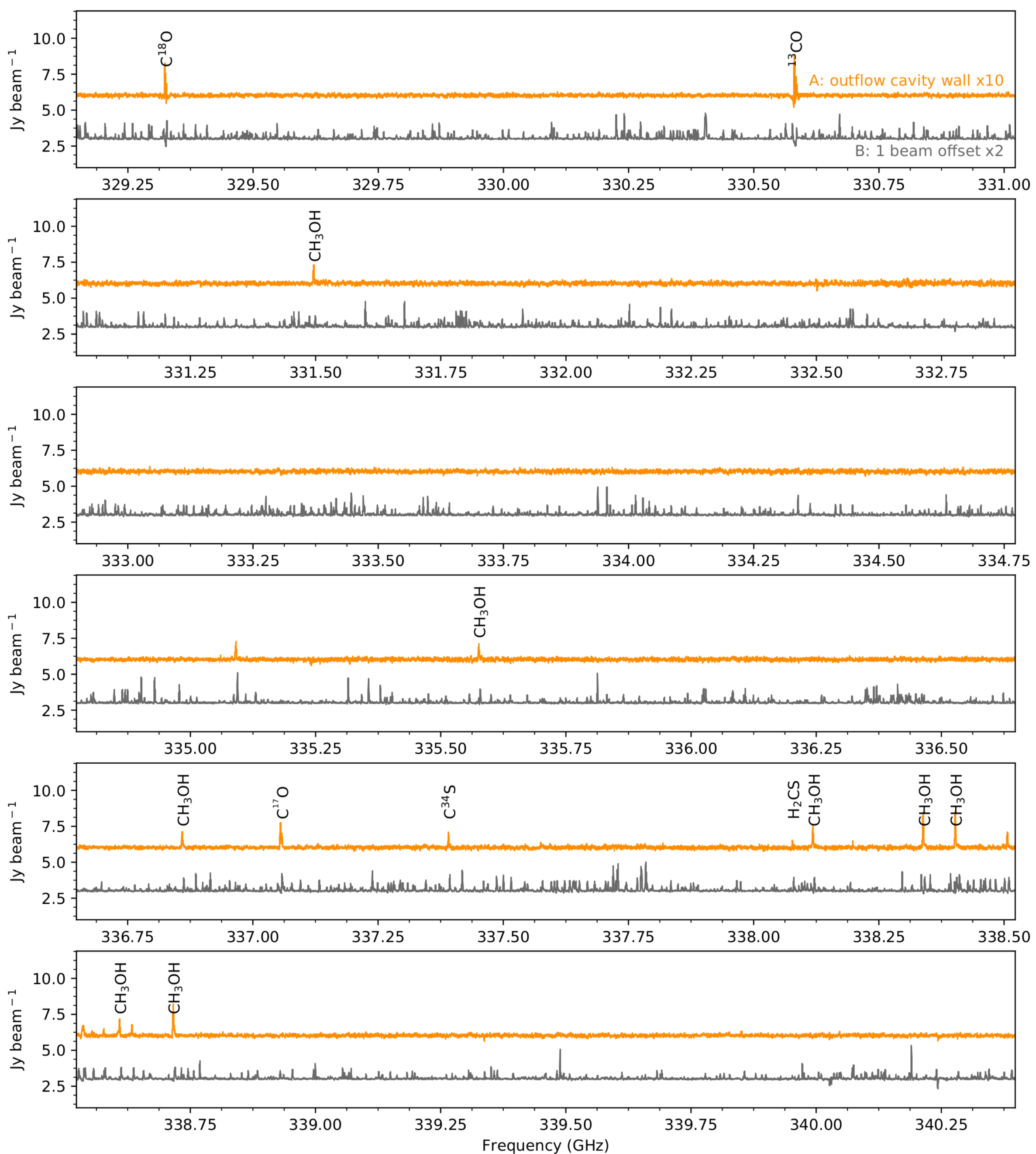}
\caption{Full spectra obtained in the PILS survey for IRAS~16293-2422. Here the frequency range 329.15 to 340.4 GHz is shown. The rest of the spectra is shown in Fig.~\ref{fig:fullspec2} and \ref{fig:fullspec3}. The spectra for the IRAS~16293-2422 B is one beam away from the source position ($\alpha_{J2000}$ = 16:32:22.581, $\delta_{J2000}$ = -24:28:32.80), note that it is multiplied by a factor of 2. The spectra for IRAS~16293-2422 A's outflow cavity wall is centered at the observed peak of \ce{c-C3H2} ($\alpha_{J2000}$ = 16:32:22.867, $\delta_{J2000}$ = -24:28:39.60), note that it is multiplied by a factor of 10 and shows very little emission other than \ce{c-C3H2}, \ce{CH3OH}, \ce{H2CS}, \ce{H2CO} and \ce{HCO+}.}
\label{fig:fullspec1}
\end{figure*}

\begin{figure*}
\centering
\includegraphics[width=0.98\textwidth]{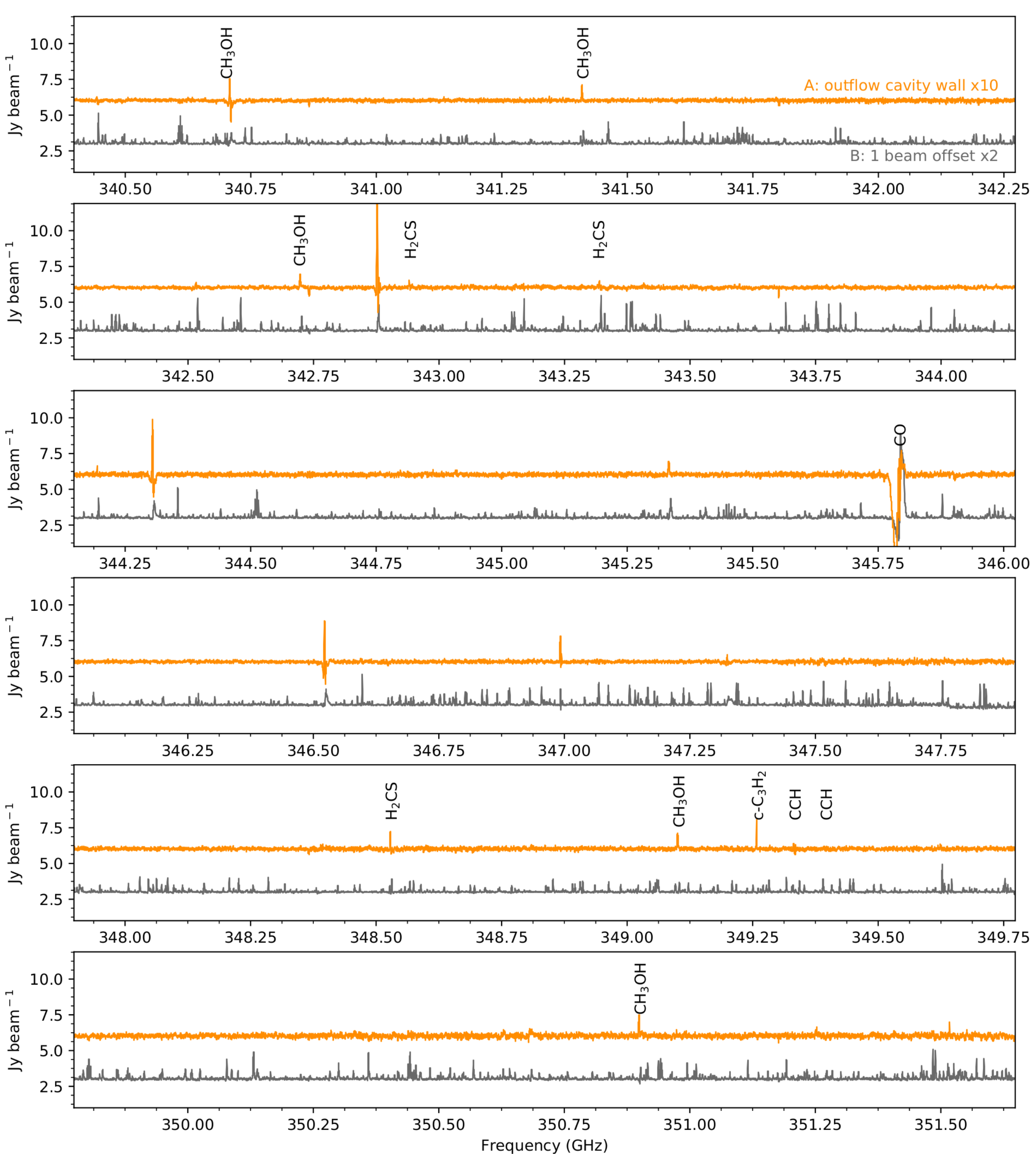}
\caption{Same as in Fig.~\ref{fig:fullspec1} but for the frequency range 340.4 to 351.65 GHz. \ce{C2H} is marked for reference.}
\label{fig:fullspec2}
\end{figure*}

\begin{figure*}
\centering
\includegraphics[width=0.98\textwidth]{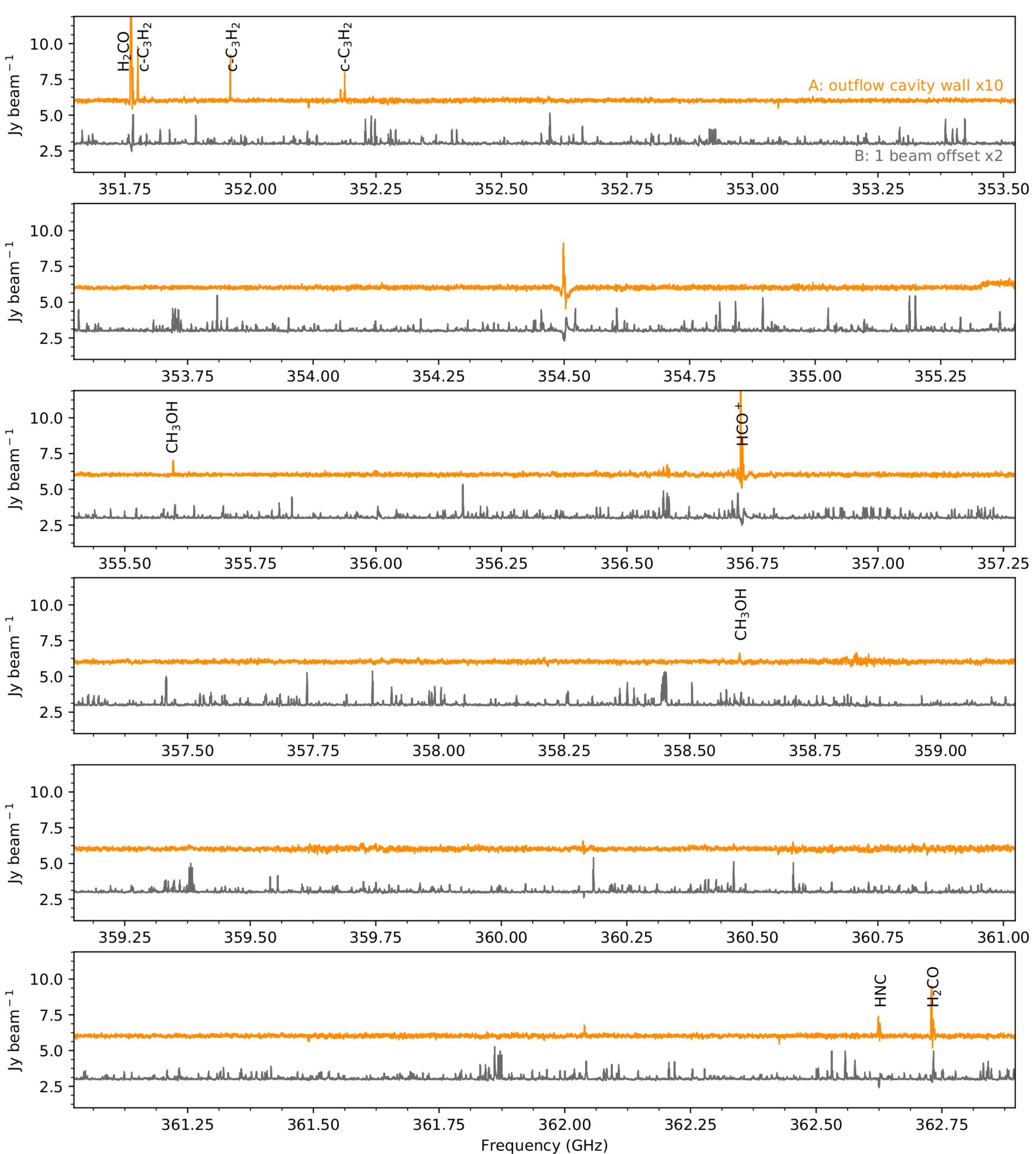}
\caption{Same as in Fig.~\ref{fig:fullspec1} but for the frequency range 351.65 to 362.9 GHz.}
\label{fig:fullspec3}
\end{figure*}

The PILS survey spectra \citep{jorgensen2016} is reproduced here for the south \ce{c-C3H2} peak position and at one beam away from the position of IRAS~16293~B .
Figures \ref{fig:fullspec1}, \ref{fig:fullspec2} and \ref{fig:fullspec3} present the full spectra for both positions.
At the south \ce{c-C3H2} peak position, the spectra is multiplied by a factor of 10 to bring out the features.
Few molecular lines are detected at this position, apart from common molecules like \ce{HCO+} and \ce{CO}, only \ce{c-C3H2} and \ce{H2CS} are detected.

\end{appendix}

\end{document}